\documentclass[final,5p,authoryear,times]{elsarticle}
\pdfoutput=1
\usepackage[colorlinks,linkcolor=blue,citecolor=blue,urlcolor=blue]{hyperref}
\usepackage{graphicx}
\usepackage{amsfonts,amsmath,amssymb}
\usepackage[utf8]{inputenc}
\usepackage{textcomp}
\usepackage{enumitem}
\usepackage{dblfloatfix} % fix problems with full-width images/tables
\usepackage{caption}
\usepackage{flushend}
\usepackage[ruled]{algorithm}
\usepackage[noend]{algpseudocode}
\usepackage[normalem]{ulem}
\alglanguage{pseudocode}

\let\origfootnote\footnote
\renewcommand{\footnote}[1]{\kern.06em\origfootnote{#1}}
\newcommand{\punctfootnote}[1]{\kern-.06em\origfootnote{#1}}

\DeclareRobustCommand{\object}[1]{%
   #1%
}

\newcommand{\tens}[1]{\mathsf{#1}}
\newcommand{\tX}{\tens{X}}
\newcommand{\tY}{\tens{Y}}
\newcommand{\tZ}{\tens{Z}}
\newcommand{\tU}{\tens{U}}
\newcommand{\tM}{\tens{M}}
\newcommand{\tA}{\tens{A}}
\newcommand{\tS}{\tens{S}}
\newcommand{\tL}{\tens{L}}
\providecommand{\norm}[1]{\left\lVert#1\right\rVert}
\renewcommand{\vec}[1]{\mathbf{#1}}
\newcommand{\x}{\vec{x}}
\newcommand{\z}{\vec{z}}
\renewcommand{\u}{\vec{u}} % won't need umlauts here
\newcommand{\prox}{\mathrm{prox}}
\newcommand{\scarlet}{{\sc scarlet}}
\journal{Astronomy \& Computing}
\begin{document}
\begin{frontmatter}

\title{\scarlet: Source separation in multi-band images by Constrained Matrix Factorization}
\author[1]{Peter Melchior}
\ead{peter.melchior@princeton.edu}
\author[1]{Fred Moolekamp}
\ead{fredem@princeton.edu}
\author[1]{Maximilian Jerdee}
\author[1]{Robert Armstrong}
\address[1]{Department of Astrophysical Sciences, Princeton University, Princeton, NJ 08544, USA}
\author[2]{Ai-Lei Sun}
\address[2]{Department of Physics \& Astronomy, Johns Hopkins University, Bloomberg Center, 3400 N. Charles St., Baltimore, MD 21218, USA}
\author[1]{James Bosch}
\author[1]{Robert Lupton}

\begin{abstract}
We present the source separation framework \scarlet\ for multi-band images, which is based on a generalization of the Non-negative Matrix Factorization to alternative and several simultaneous constraints.
Our approach describes the observed scene as a mixture of components with compact spatial support and uniform spectra over their support. 
We present the algorithm to perform the matrix factorization and introduce constraints that are useful for optical images of stars and distinct stellar populations in galaxies, in particular symmetry and monotonicity with respect to the source peak position.
We also derive the treatment of correlated noise and convolutions with band-dependent point spread functions, rendering our approach applicable to coadded images observed under variable seeing conditions.
\scarlet\ thus yields a PSF-matched photometry measurement with an optimally chosen weight function given by the mean morphology in all available bands.
We demonstrate the performance of \scarlet\ for deblending crowded extragalactic scenes and on an AGN jet -- host  galaxy separation problem in deep 5-band imaging from the Hyper Suprime-Cam Stategic Survey Program.
Using simulations with prominent crowding we show that \scarlet\ yields superior results to the HSC-SDSS deblender for the recovery of total fluxes, colors, and morphologies.
Due to its non-parametric nature, a conceptual limitation of \scarlet\ is its sensitivity to undetected sources or multiple stellar population within detected sources, but an iterative strategy that adds components at the location of significant residuals appears promising.
The code is implemented in {\sf Python} with {\sf C++} extensions and is available at \url{https://github.com/fred3m/scarlet}.
\end{abstract}

\begin{keyword}
%% keywords here, in the form: keyword \sep keyword
% Astronomy keywords from: http://iopscience.iop.org/article/10.1086/318271/fulltext
% Computing keywords from: http://www.acm.org/about/class
 methods: data analysis \sep  techniques: image processing \sep galaxies: structure \sep Non-negative matrix factorization
%% MSC codes here, in the form: \MSC code \sep code
%% or \MSC[2008] code \sep code (2000 is the default)

\end{keyword}

\end{frontmatter}

\section{Introduction}
\label{sec:intro}

Modern astronomical wide-field surveys cover large areas of the sky at ever increasing depths, revealing more objects and low-surface-brightness features of extended objects that were previously too faint to detect.
These gains drive investigations into galactic, extragalactic and cosmological phenomena at an unprecedented level of detail and statistical power.
On the other hand, because of the enhanced sensitivity, a larger fraction of the observed area is associated with detectable objects, thereby increasing the chance that multiple objects overlap.
This so-called ``blending'' constitutes a major concern for the analysis of existing and upcoming surveys, especially those that observe from the ground.

The majority of methods for measuring the properties of celestial objects assume that every object can be considered isolated.
If that assumption holds, well-defined and accurate measurements of the flux, position, shape, and morphology can routinely be made with methods that are either based on moments of the light distribution within some aperture or on parametric fits to the images. 

However, the notion of isolated objects is becoming increasingly obsolete.
With a limiting magnitude of $i\approx24$, DES\footnote{\url{https://www.darkenergysurvey.org}} \citep{DES2016} finds that 30\% of galaxies suitable for weak-lensing measurements are affected by blending \citep{Samuroff17}. 
For HSC\footnote{\url{http://hsc.mtk.nao.ac.jp/ssp/}} \citep{Aihara18.1}, whose Wide survey has a limiting magnitude of $i\approx26$, \citet{Bosch17} find that 58\% of measured objects are in blended groups, a dramatic increase despite a substantially better average seeing than DES.
LSST\footnote{\url{http://lsst.org}} \citep{Ivezic08.1} expects to reach $i\approx27$ after 10 years of operations , and it is estimated that 63\% of observed galaxies will have S\'ersic model photometry that is altered by more than 2\% due to the presence neighbors  (Sanchez et al., in prep.).
Even more problematically, in a comparison study of HST and Subaru imaging of a galaxy cluster field, where the Subaru data had similar seeing and depth so as to serve as a proxy for LSST, \citet{Dawson16} found that 14\% of observed galaxies are blended but not recognized as such in the ground-based images.
Slightly larger numbers for unrecognized blends are found for HSC in a study that inserted fake objects into  real survey images to infer how many of them could be recovered (Murata et al., in prep.).

This lack of separability between objects necessitates the employment of techniques that analyze entire scenes with overlapping objects.
For direct measurements, such as moments within apertures, there is no accurate way to correct for the excess light from overlapping objects because such a correction depends e.g. on radial profiles of the objects involved, which cannot be determined well for blends.
As a consequence, multiple objects need to be \emph{modeled} either iteratively (by masking all but one) or simultaneously, often requiring sophisticated and fine-tuned schemes to prevent unstable or physically implausible solutions \citep[e.g.][]{Barden12.1, Drlica-Wagner17.1}.
Any such scheme is suitable to extract and separate the objects in celestial scenes, i.e. to ``deblend'' those scenes.
Differences between the schemes include the propagation of errors, which conceptually favors simultaneous approaches, and whether the desired measurements are generated directly from deblender models or by passing them on to established measurement algorithms for isolated objects.

Traditional deblending approaches in astronomy are achromatic, i.e. they employ information from only a single image.
{\sc SExtractor} detects blending by thresholding an image at a range of intensity levels and searching for sets of pixels that are connected at a lower threshold but split into several connected regions at a higher one \citep{Bertin96.1}.
As a consequence of the splitting approach, the association of pixels to objects is unique and exclusive, i.e. in the internal representation of blended objects they do not overlap.
This unrealistic notion has necessitated mitigation strategies or fine-tuning to prevent ``over-deblending'' of larger galaxies caused by smaller and fainter companions or interlopers \citep[e.g.][]{Rix04.1}.

The deblender in the SDSS {\it Photo} pipeline (Lupton, in prep.) does allow for overlap between nearby objects and, consequently, needs to estimate the portion of any pixel's flux that is due to each object.
It uses a two-step approach, in which first a template is constructed for each object based on the requirement that pixel values symmetrically across the object's peak pixel be identical; they are generally not, so the minimum pixel value of those two pixels is adopted for both.
Then, the original image values are projected onto those templates, associating each pixel's flux to different objects in proportion to the amplitude of the respective templates.
%\begin{enumerate}
%    \item A template is constructed for each object based on the requirement that pixel values on opposite sides an object's peak pixel are identical. This is not generally true, but it is very nearly true for most elliptical and spiral galaxies and allows the lack of flux on the unblended side of an object to constrain the shape on the other side (which may be blended with a neighbor). So each pixel and it's symmetric partner adopt the minimum pixel value of the pair.
%    \item The original image values are projected onto these symmetric templates, associating each pixel's flux to different objects in proportion to the amplitude of their respective templates.
%\end{enumerate}
Despite very few assumptions, the method mostly separates sources into physically plausible objects but struggles with situations where a central object is symmetrically surrounded by neighbors, for instance a blend with three peaks in a row (Lupton, in prep.).

\begin{figure*}[ht!]
    \includegraphics[width=.332\linewidth]{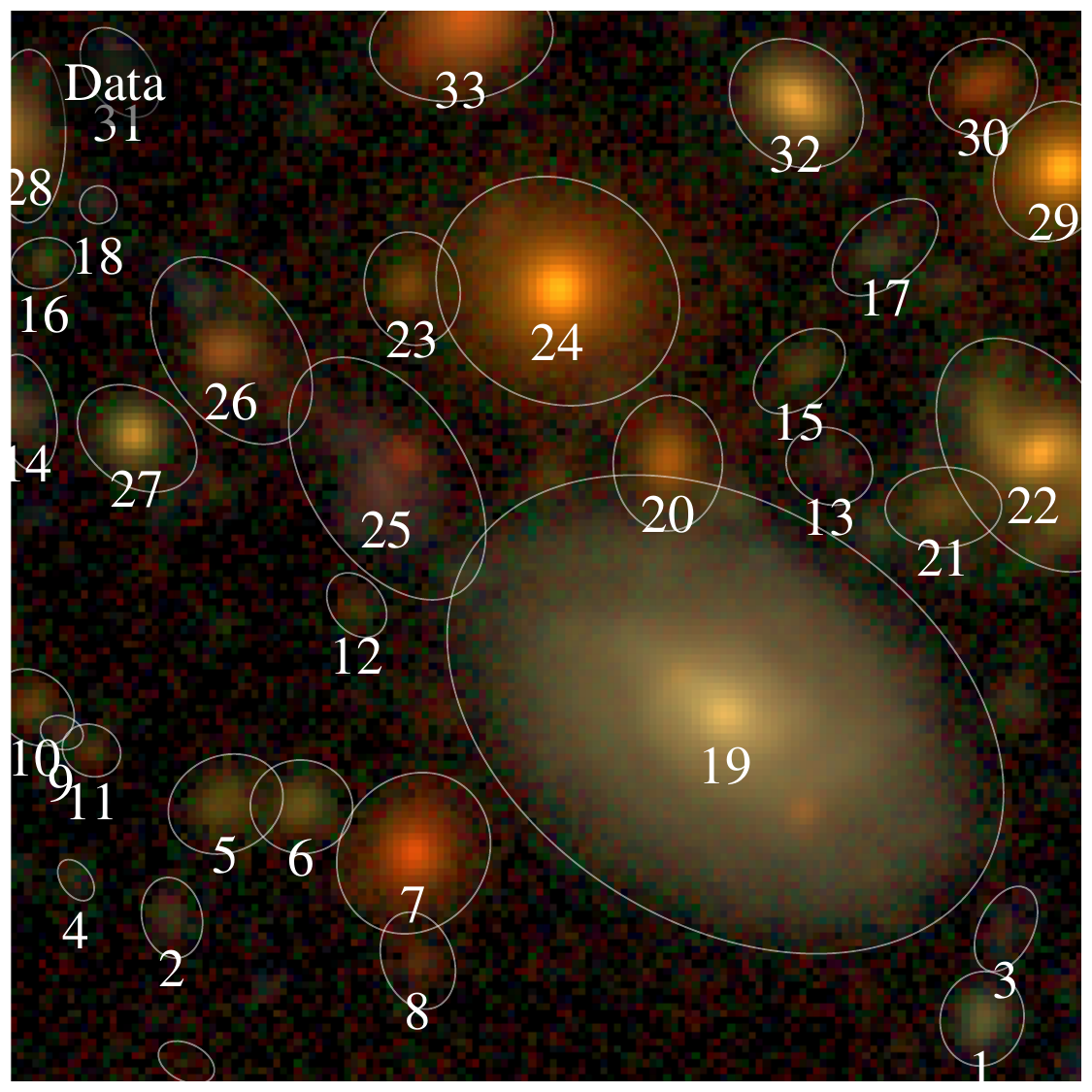}
    \includegraphics[width=.332\linewidth]{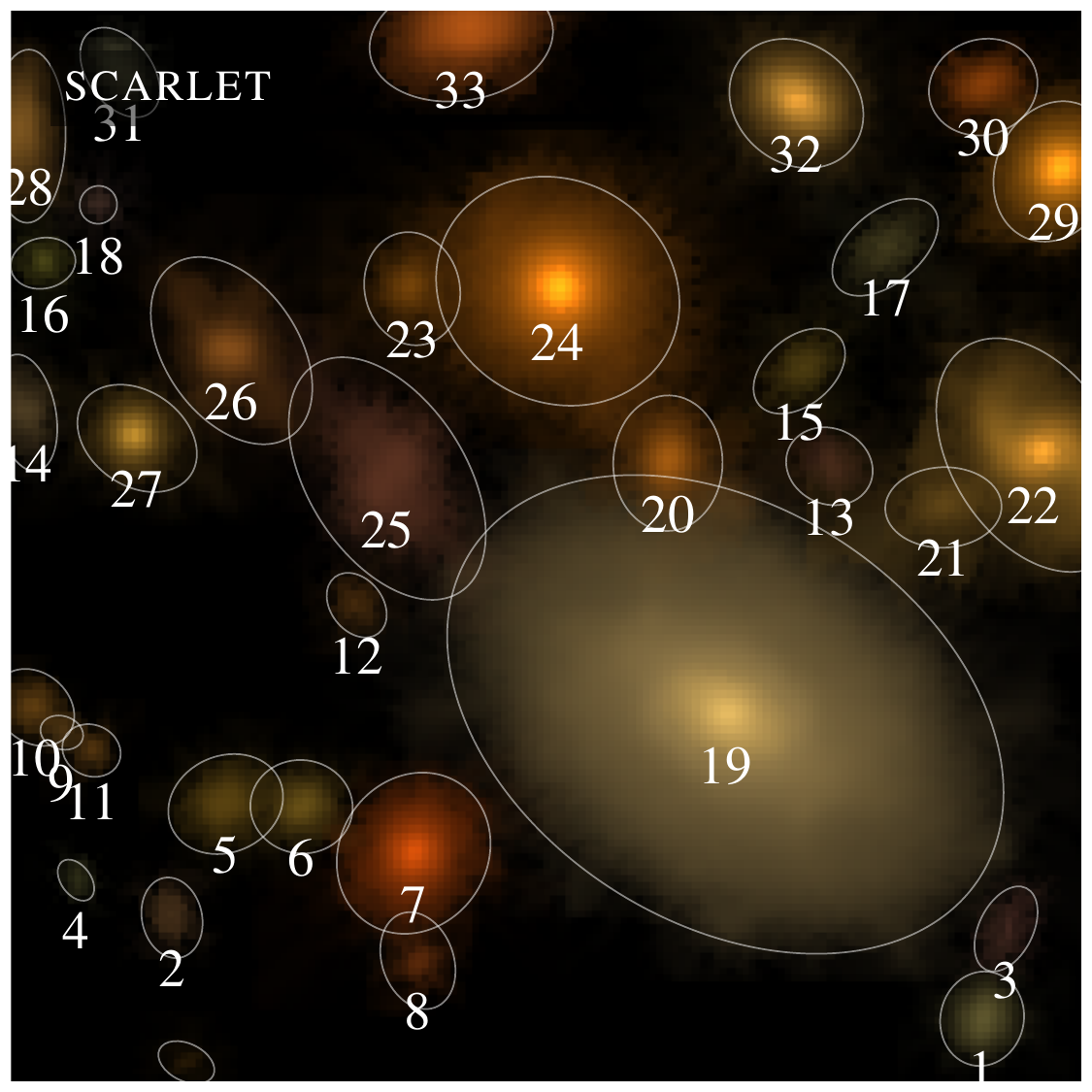}
    \includegraphics[width=.332\linewidth]{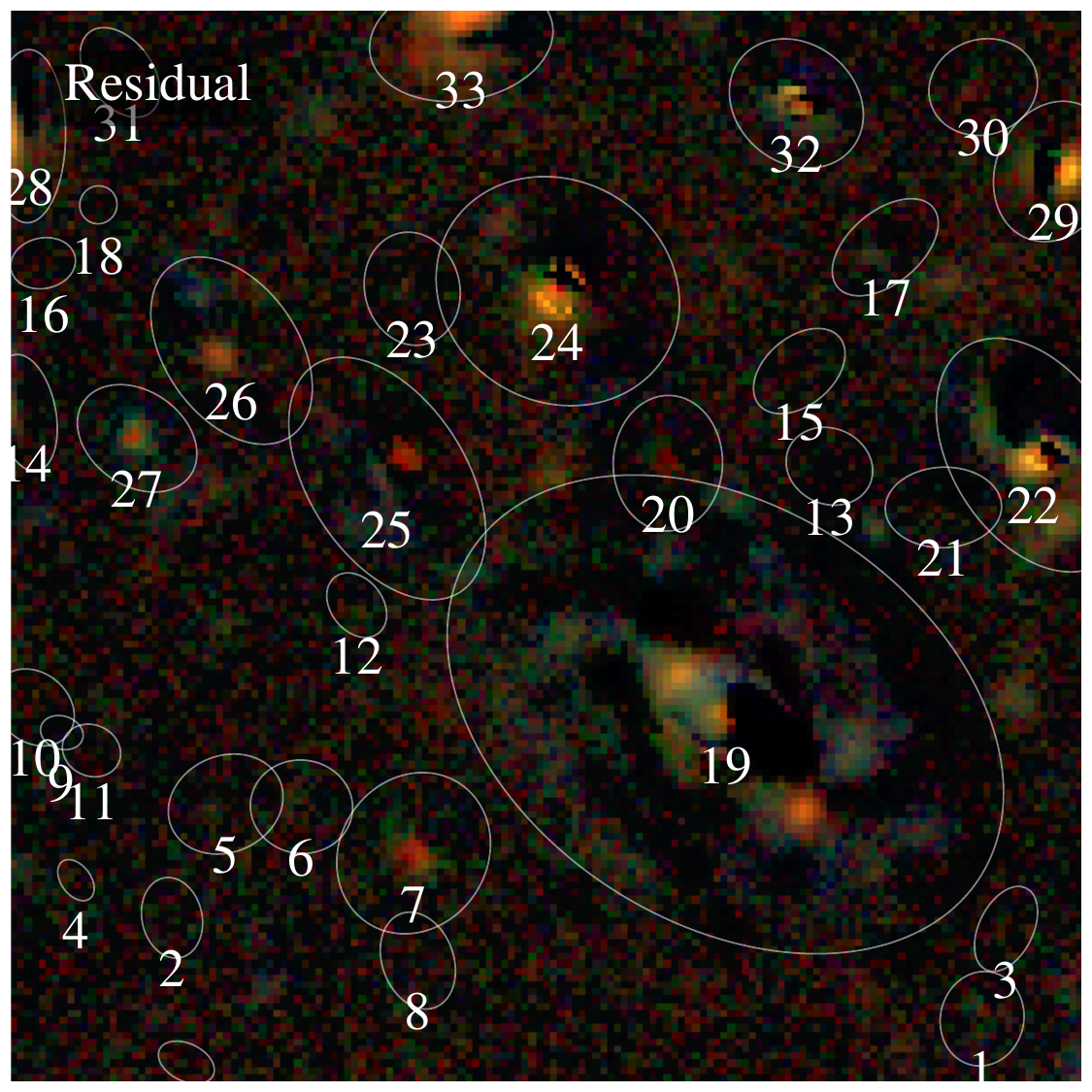}
     \caption{\emph{Left:} False-color image of $grizy$ coadd images $\tY$ from the HSC UltraDeep COSMOS data release, shown with an arcsinh stretch. The scene spans $25\times25$ arcsec$^2$.  Object detections and ellipse fits are performed by {\sc SExtractor} on the detection image (sum of the five coadds). \emph{Center:} \scarlet\ model $\tA\tS$ of the scene with single-component sources for each detection. \emph{Right:} The residual $\tY-\tA\tS$ reveals the presence of additional sources or color variations within detected sources.}
     \label{fig:hsc}
\end{figure*}

More recently, \citet{Zhang15.1} and \citet{Connor17.1} proposed variants of inpainting techniques, where the relevant pixels of blended objects are replaced by an estimate of a local variable background, working inwards from an initially defined outline.
The portion of the pixel flux above the background estimate is attributed to the respective object.
These approaches implicitly account for blending by assuming that the background captures the flux contributions from neighboring objects.
This is particularly useful when recovering small objects in multi-scale blending situations like the cores of galaxy clusters and  removing them from the scene so that large objects can be measured separately.

While effective in many cases, all of the deblending schemes outlined above employ heuristic arguments for how to separate overlapping sources.
They also perform the pixel--object association sequentially, one object at a time, thus losing the advantages of a simultaneous solution, for instance the ability to explore the degeneracies that arise because the objects are not isolated. 
However, it is our opinion that the biggest limitation stems from the restriction to a single image, and therefore a single filter band, while most modern surveys observe the sky in several filters. 
A visual inspection of multi-band images clearly suggests that color can serve as a powerful discriminator between different objects, even with severe overlap (see \autoref{fig:hsc}).

{\sc MuSCADeT} \citep{Joseph16.1} addresses both limitations by building a joint model of multi-band image data.
As their model is non-parametric, the number of degrees of freedom is large, which leads to many possible degeneracies in the solutions, so they demand that the spatial distribution of each source be sparse in the starlet (a form of isotropic wavelet) domain.  
The resulting solutions extract preferentially compact features down to the noise level, using a set of previously identified colors for each feature.
For applications to wide-field multi-band data, we cannot generally assume to a priori identify the color of an object in the scene because there might not be a single pixel whose color is uncontaminated by other objects.
We therefore seek the ability to update both spectral and morphological characteristics of the objects.
While sharing noticeable similarities with {\sc MuSCADeT} regarding the use of a non-parametric constrained morphological model, one can consider our approach an extension that also updates the source spectra as well as a generalization that allows an arbitrary number of constraints to be placed on each source.

The outline of the paper is as follows: We introduce our approach, dubbed \scarlet, in \autoref{sec:method}, demonstrate its performance on real data and simulations in \autoref{sec:results}, and conclude in \autoref{sec:conclusions}. 

\section{Methodology}
\label{sec:method}

We base our deblender on the assumption that astronomical scenes are superpositions of multiple components, each with
\begin{enumerate}[label=\arabic*., ref=\arabic*., leftmargin=*, partopsep=0em, itemsep=0em ]
\item a spatially compact support and
\item a constant spectrum over that support.
\end{enumerate}
For stellar fields this is obviously true, but even for galaxies, especially marginally resolved ones, which constitute the vast majority of galaxies in deep surveys, the assumption is appropriate.
In addition, even large, extended galaxies can be thought of as conglomerations of components (e.g. bulges, discs, bars, star forming regions) for which the assumptions above hold at least approximately.%
\footnote{In the literature the terms ``source separation'' and ``component separation'' are often used interchangeably. To better reflect the hierarchical nature of astrophysical scenes, we will define the term ``Source'' as a collection of co-centered components that belong to the same astrophysical object.}
For instance, the popular bulge-disc decomposition for galaxies \citep[e.g.][]{MacArthur13.1} is justified by this interpretation.
We note that the assumption of linear superposition implies that components do not interact, which is correct only for transparent emitters.
Absorption, e.g. by dust in the galaxy, can be approximated by allowing negative values in the source spectrum, but substantial opacities cannot fully be modeled because the effect depends on the amount of absorbing material \emph{and} the intensity of the background radiation.

The assumptions above appear to lend themselves to a parametric modeling framework, where one assumes to know the shapes of the components and potentially their intrinsic spectrum, exploiting quite tight relations between colors and morphologies exhibited by galaxies in the late universe \citep[e.g.][]{Conselice98.1,Ball08.1}.
While drastically reducing the number of optimization parameters, we are critical of this approach for two reasons:
First, in the translation of an intrinsic, restframe spectrum to the observed broadband colors one needs to take the galaxy's redshift into account, which is equivalent to estimating a photometric redshift as part of the deblending process.
If the redshifting prescription is incorrect, e.g. because of a limited library of spectra or the evolution of those spectra with redshift,
it would affect the properties of the deblended components---not only their recovered spectra but also their shapes.
Second, at the stage in the analysis pipelines of large astronomical surveys where we envision the deblender to operate, it will not necessarily be established what kind of sources are in the scene; in other words, a suitable parameterization is probably not known.
This is most evident when looking at the star-galaxy distinction: for stars, three parameters are sufficient (two centroid coordinates and one amplitude), while even simple galaxy models need at least one more parameter (the size).
Model-fitting under those conditions can be done by transdimensional sampling \citep[e.g.][]{Green95.1}, but the computational costs are likely too high for large-volume data sets.

Because of these concerns, we seek to characterize the scenes without making questionable astrophysical assumptions, which means describing colors in the observed frame and morphologies in the free-form space of image pixels.

\subsection{Non-negative Matrix Factorization}
\label{sec:nmf}

We assume that an astronomical scene $\tY$ that we seek to analyze is organized in the form of a multi-band image cube of aligned images in $B$ bands, each of which is suitably flattened to have a total number of $N$ pixels.
Our previous assumptions give rise to a multi-band model $\tM$ as a sum of a finite number of components $K$,
\begin{equation}
\label{eq:model}
\tM = \sum_{k=1}^K \tA^\top_k \times \tS_k = \tA\tS,
\end{equation}
where $\tA_k\in \mathbb{R}^B$ is the amplitude of component $k$ across all bands, i.e. its spectral energy distribution (SED), and $\tS_k\in\mathbb{R}^N$ is the spatial shape of that component. By arranging the $\tA_k$ as columns of $\tA\in\mathbb{R}^{B\times K}$ and $\tS_k$ as the rows of $\tS\in\mathbb{R}^{K\times N}$, we have a model $\tM$ that is given by the product of two matrix factors.

With a homoscedastic Gaussian error model, which is appropriate for most extragalactic images in the optical, the likelihood function is
\begin{equation}
\label{eq:simple_f}
f(\tA,\tS) = \frac{1}{2}\norm{\tY-\tA\tS}_2^2
\end{equation}
where $\norm{.}_2$ denotes the element-wise $L_2$ (Frobenius) norm.%
\footnote{The objective function in \autoref{eq:simple_f} is insufficient for dealing with realistic multi-band data, which generally exhibit correlated noise and correlated signals, the former from warping the images to rectify any astrometric distortion, the latter because of the blurring from the point spread function.
We will work out in \autoref{sec:weights} and \autoref{sec:psf} how to deal with both effects.}
In its simplest form, the Non-negative Matrix Factorization \citep[NMF][]{Paatero1994} then amounts to fitting $\tA$ and $\tS$ such that they minimize $f$ and obey the non-negativity constraint, which is given by the indicator function of the  set of non-negative matrix elements:
\begin{equation}
\label{eq:gplus}
g_+(\tens{X}) = \begin{cases}
0 & \text{if}\ \tens{X}_{mn} \geq 0\ \forall m,n\\
\infty & \text{else}.
\end{cases}
\end{equation}
In other words, one seeks to minimize $f(\tA,\tS) + g_+(\tA) + g_+(\tS)$.
The classical way of solving this NMF problem is known as ``multiplicative updates'' \citep{Lee2001}, which suffers from poor convergence if the constraints strongly work against the minimum of the objective function.
Moreover, in its simplest form the NMF is hampered by a degeneracy that stems from the transformation $(\tA,\tS)\rightarrow(\tA\tens{Q},\tens{Q}^{-1}\tS)$ with an arbitrary invertible matrix $\tens{Q}$, which means power can be shifted from $\tA$ to $\tS$ or vice versa, hampering convergence to a well-defined solution.
To compensate one can introduce another constraint on the norms of either $\tA$ or $\tS$.
The most natural normalization for astronomical scenes is to require that each component's SED sums up to unity.
We therefore seek to minimize $f(\tA,\tS) + g_+(\tA) + g_\mathrm{unity}(\tA) + g_+(\tS)$, where $g_\mathrm{unity}$ is another indicator function of the set of matrices in which the columns are properly normalized.

We do so by using a two-block variant of the ``proximal gradient method''.
In short, a closed smooth proper convex function $f(\vec{x})$ under a convex constraint function $g(\vec{x})$ with $\vec{x}\in \mathbb{R}^n$ can be minimized with the following sequence:
\begin{equation}
\label{eq:pgm}
\vec{x}^{\mathrm{it}+1} \leftarrow \prox_{\lambda^\mathrm{it} g} \left(\vec{x}^\mathrm{it} - \lambda^\mathrm{it} \nabla f(\vec{x}^\mathrm{it})\right).
\end{equation}
The step size $\lambda$ must be chosen from $(0,2/L]$, where $L$ is the Lipschitz constant of $\nabla f$.
The function $g$ is not directly evaluated but accessed through its \emph{proximal operator},
\begin{equation}
\label{eq:proximal}
\prox_{\lambda g}\left(\vec{x}\right)\equiv\underset{\vec{v}}{\textrm{argmin}}\left\lbrace g\left(\vec{v}\right)+\frac{1}{2\lambda}\norm{\vec{x}-\vec{v}}_{2}^{2}\right\rbrace.
\end{equation}
The proximal operator of $g$ effectively hides the infinite values and the non-differentiability of its underlying function.
For instance, the proximal operator of the indicator function $I_\mathcal{C}$ of a non-empty set $\mathcal{C}$ is the Euclidean projection operator onto the set---i.e. it yields $\vec{v}\in\mathcal{C}$ that is nearest to $\x$---and the scaling parameter $\lambda$ is irrelevant.
In particular, 
\begin{equation}
\label{eq:prox_plus_unity}
\begin{split}
&\prox_{+}(\vec{x}) = \left(x_1^+, \dots, x_n^+\right)\ \text{with}\ x_i^+ \equiv \text{max}(0,x_i)\\
&\prox_{\mathrm{unity}}(\vec{x}) = \norm{\vec{x}}_1^{-1} \vec{x},\\
\end{split}
\end{equation}
For many other classes of constraint functions proximal operators have analytic forms, which means that the minimization of \autoref{eq:proximal} does not have to be carried out.
For more details on proximal methods, we refer the reader to \citet{Combettes2009} and \citet{Parikh2014}.

As long as there is only one constraint per variable, we can interleave update steps of the form of \autoref{eq:pgm} that minimize $f(\tA\,|\,\tS)$ and those that minimize $f(\tS\,|\,\tA)$, effectively a proximal implementation of a coordinate descent method, as shown in \autoref{alg:pgm}. Note that we merged $g_+(\tA) + g_\mathrm{unity}(\tA)$ into one proximal operator $\prox_{\mathrm{unity}+}\equiv \prox_+\left(\prox_\mathrm{unity}\right)$, which can be done because the two constraints commute, i.e. they do not interfere.
This reduces the problem to calculating the gradients
\begin{equation}
\label{eq:gradients}
\begin{split}
&\nabla_\tA f(\tA,\tS) = -(\tY-\tA\tS)\,\tS^\top\\
&\nabla_\tS f(\tA,\tS) = -\tA^\top(\tY-\tA\tS)\\
\end{split}
\end{equation}
and their Lipschitz constants
\begin{equation}
\label{eq:lipschitz}
L_\tA = \norm{\tS \tS^\top}_s\ \text{and}\ L_\tS = \norm{\tA^\top \tA}_s,
\end{equation}
where $\norm{.}_s$ refers to the spectral norm. 
The sequence in \autoref{alg:pgm} approaches a fixed point that is also a local minimum of $f$ and can therefore be terminated whenever the change between successive iterations falls below a user-defined threshold.

This approach is similar to the one of \citet{Rapin2013} in that the minimization is performed in two blocks and that constraints are expressed by their proximal operators, but we only perform one gradient step instead of seeking the exact solution for every block in \emph{every iteration}.
This reduces the number of gradient and constraint evaluations while still yielding a sequence that is guaranteed to converge \citep{Lin2007,Xu2013}.

\begin{algorithm}[t]
\caption{Proximal Gradient NMF\newline
{\small Two matrix factors $\tA$ and $\tS$ are independently updated with a sequence of gradient steps followed by projections to obey the constraints.}}
\label{alg:pgm}
\begin{algorithmic}[1]
\Procedure{NMF-PGM}{$\tA^0, \tS^0$}
\For{$\text{it} = 0,1,\dots$}
\State $\lambda_{\tA}^{\mathrm{it}} \gets \norm{\tS^\mathrm{it} \tS^{\mathrm{it}\top}}_s^{-1}$
\State $\tA^{\mathrm{it}+1} \gets \prox_{\mathrm{unity}+}\left(\tA^{\mathrm{it}} - \lambda_{\tA}^{\mathrm{it}} \nabla_\tA f(\tA^\mathrm{it},\tS^\mathrm{it})\right)$
\State $\lambda_{\tS}^{\mathrm{it}} \gets \norm{\tA^{\mathrm{it}+1\,\top} \tA^{\mathrm{it}+1}}_s^{-1}$
\State $\tS^{\mathrm{it}+1} \gets \prox_+\left(\tS^{\mathrm{it}} - \lambda_{\tS}^{\mathrm{it}} \nabla_\tS f(\tA^{\mathrm{it}+1},\tS^\mathrm{it})\right)$
\EndFor
\EndProcedure
\end{algorithmic}
\end{algorithm} 

\subsection{Constrained Matrix Factorization}

For many practical applications we want to be able to impose additional constraints besides (or instead of)  non-negativity.
We will give a list of alternative constraints in \autoref{sec:constraints}, but it is useful to think of constraints more generally as logarithms of priors on the solution.
In addition, we are often forced to impose multiple simultaneous constraints to deal with the large number of unknowns in \autoref{eq:model}, especially when the number of components $K$ is larger than the number of bands $B$.
To reflect these more general conditions, we favor the term ``Constrained Matrix Factorization'' (CMF) instead of NMF.

With the direct projection method of \autoref{eq:pgm}, we can only accommodate any one proximal operator per variable, even though non-interacting constraints can be daisy-chained to act like a single one, as we have done with $\prox_{\mathrm{unity}+}$.
One could generalize this approach by the method of Alternating Projections, also known as Projection onto Convex Sets \citep{Bauschke1996}, but that still requires that the proximal operators act directly on the variables $\tA$ and $\tS$.
Often, it is beneficial to express constraints in a transformed domain, for instance demand that the solution be sparse in that domain \citep[e.g.][]{Rapin2014}.
We therefore prefer the Alternating Direction Method of Multipliers \citep[ADMM,][]{Gabay1976, Glowinski1975, Eckstein1992, Boyd2011}.
In its basic form it seeks to
\begin{equation}
\label{eq:ADMM}
\underset{\vec{x}}{\text{minimize\:}}f(\vec{x})+g\left(\tens{L}\vec{x}\right)
\end{equation}
which allows us to introduce an arbitrary matrix $\tens{L}$.
For this work $\tens{L}$ will be a gradient or symmetry operator, but it could as well be a Fourier or wavelet transform, etc.
The previous equation can be re-written in ``consensus form''
\begin{equation}
\begin{array}{cc}
\textrm{{minimize}} & f\left(\vec{x}\right)+g\left(\vec{z}\right)\\
\textrm{{subject\,to}} & \tens{L}\vec{x}-\vec{z}=0,
\end{array}
\end{equation}
which suggests an approach that splits the optimization into two separate tasks: one that minimizes $f$ and another that satisfies $g$ by introducing the auxiliary variable $\vec{z}$ \citep{Douglas1956}.
The updating scheme also introduces the dual variable $\vec{u}$, which connects the two tasks, creating the following sequence:
\begin{equation}
\label{eq:admm}
\begin{array}{ccl}
\x^{\mathrm{it}+1} & \leftarrow & \prox_{\lambda f}\left(\x^\mathrm{it}-\frac{\lambda}{\rho} \tL^{\top}\left(\tL\x^\mathrm{it}-\z^\mathrm{it}+\u^\mathrm{it}\right)\right)\\
\z^{\mathrm{it}+1} & \leftarrow & \prox_{\rho g}\left(\tL\x^{\mathrm{it}+1}+\u^\mathrm{it}\right)\\
\u^{\mathrm{it}+1} & \leftarrow & \u^\mathrm{it}+\tL\x^{\mathrm{it}+1}-\z^{\mathrm{it}+1},
\end{array}
\end{equation}
where the scaling parameters need to satisfy $0<\lambda\leq\rho/||\tL||_\mathrm{s}^2$.%
\footnote{For the $\x$-update, we use a linearized form of the ADMM, which is known as \emph{split inexact Uzawa method} \citep[e.g.][]{Esser2010, Parikh2014}.}
Two aspects of this sequence are remarkable. 
The function $f$ is only accessed through its proximal operator, which means it does not need to be differentiable.
In our case, $f$ is differentiable, and we can consider
\begin{equation}
\label{eq:prox_f}
\prox_{\lambda f}(\x) \equiv \vec{x} - \lambda \nabla f(\x)
\end{equation}
the corresponding proximal operator of the first-order approximation to $f$.%
\footnote{In fact, we can treat the entire projected gradient step of \autoref{eq:pgm} as a first-order approximation of $f + g$, which means we can satisfy a single constraint in the direct domain, such as non-negativity, without having to invoke the auxiliary variables $\z$ and $\u$.}
Second, the linear operator $\tL$ does not need to be invertible; it and its transpose are only used for dot products, rendering the updates particularly efficient with sparse matrices. 

In \citet{Moolekamp2017}, we have demonstrated that the ADMM can be extended to multiple variables of a function $f$ that is convex in each of those variables by constructing a block-updating sequence analogous to \autoref{alg:pgm}.
For the CMF, $\tA$ and $\tS$ (which we will call $\tX_1$ and $\tX_2$ and enumerate with the index $j$ for the remainder of this subsection) thus have their own set of ADMM auxiliary variables $\tZ_1, \tU_1,\tZ_2, \tU_2$. 
In addition, we have introduced a method that allows each $\tX_j$ to be subject to multiple simultaneous constraints, one pair ($\tZ_{ji}, \tU_{ji})$ for each constraint $g_{ji}(\tX_j)$ with $\ i\in\lbrace 1,\dots,M_j \rbrace$.
Combined, our new method, dubbed \scarlet, provides the flexibility we seek for deblending astronomical scenes via the CMF.
It will 
\begin{equation}
\begin{array}{cl}
\underset{\tX_1, \tX_2}{\text{minimize}} & 
f\left(\tX_1, \tX_2\right) + \sum_{j=1}^2 \sum_{i=1}^{M_j}g_{ji}\left(\tZ_{ji}\right)\\
\textrm{subject\,to} & \tL_{ji}\tX_j-\tZ_{ji}=0\ \ \forall j\in\lbrace1,2\rbrace\ \text{and}\ i\in\lbrace 1,\dots,M_j \rbrace
\end{array}
\end{equation}
through the Block-Simultaneous Method of Multipliers \citep[bSDMM][]{Moolekamp2017}.

The core working principle of \scarlet\ is given in \autoref{alg:bsdmm}.
The ADMM updates of \autoref{eq:admm} are performed in lines 10--13, using the gradient updates from \autoref{eq:gradients} in the form of \autoref{eq:prox_f} in line 10.
The step sizes $\lambda_j$ in line 8 are given as before by $1/L_j$ with the Lipschitz constants from \autoref{eq:lipschitz}.
It is worthwhile pointing out that in the limit of all constraints being satisfied, $\tZ=\tX$ and $\tU=\tens{0}$, in which case \autoref{alg:bsdmm} is equivalent to \autoref{alg:pgm}.

Following \citet{Boyd2011}, we implement stopping criteria (on line 14) based on primal residual $\tens{P}^{\mathrm{it}+1}=\tL \tX^{\mathrm{it}+1}-\tZ^{\mathrm{it}+1}$ and the dual residual $\tens{D}^{\mathrm{it}+1}=\frac{1}{\rho}\tL^\top\left(\tZ^{\mathrm{it}+1}-\tZ^{\mathrm{it}}\right)$.
These residuals describe how close the current $\tX$ is to satisfying the constraint in the transformed domain and how much $\tZ$ changes, respectively.
To assess primal and dual feasibility, we require
\begin{equation}
\label{eq:residuals}
\begin{split}
&\norm{\tens{P}^{\mathrm{it}+1}}_2 \leq \epsilon^\mathrm{pri} \equiv \sqrt{p}\, \epsilon^\mathrm{abs} + \epsilon^\mathrm{rel} \max\left\lbrace \norm{\tens{L} \tX^{\mathrm{it}+1}}_2, \norm{\tZ^{\mathrm{it}+1}}_2\right\rbrace\ \ \text{and}\\
&\norm{\tens{D}^{\mathrm{it}+1}}_2 \leq \epsilon^\mathrm{dual} \equiv \sqrt{n}\, \epsilon^\mathrm{abs} + \epsilon^\mathrm{rel} / \rho\, \norm{\tens{L}^\top \tU^{\mathrm{it}+1}}_2,
\end{split}
\end{equation}
where $p$ and $n$ are the number of elements in $\tZ$ and $\tX$, respectively.
The error thresholds $\epsilon^\mathrm{abs}$ and $\epsilon^\mathrm{rel}$ can be set at suitable values, depending on the precision and runtime constraints of the application.

The procedure outlined in \autoref{alg:bsdmm} yields point estimates of the matrix factors.
Uncertainties on each factor are currently calculated by linear error propagation based on \autoref{eq:simple_f}.
While those uncertainties provide estimates of the significance of each element of the matrix factors, they do not properly capture the interdependence of the model components or the effects of the constraints on the degeneracies between them.
Fully accurate uncertainties are difficult to estimate because of the non-differentiability of the constraint functions.
We plan to investigate ways to better characterize the uncertainties of the \scarlet\ model in forthcoming work.

This concludes our presentation of the general approach we have adopted in \scarlet.
A reader who is mostly interested in the astrophysical applications can safely jump to \autoref{sec:results} and skip details of the formalism below.

\begin{algorithm}[t]
\caption{Block-SDMM Constrained Matrix Factorization\newline
{\small The matrix factors $\tA$ and $\tS$ are treated as two blocks $X_j$ with $j\in\lbrace1,2\rbrace$ and independently updated with a sequence of gradient steps $\prox_{\lambda_j f,j}$ and jointly optimized with an arbitrary number of constraint functions $g_{ji}$ with $i\in\lbrace1,\dots,M_j\rbrace$.
All constraints are expressed as proximal operators $\prox_{\rho_{ji} g_{ji}}$ and include transformation matrices $\tL_{ji}$ (cf. \autoref{eq:admm} for the generic ADMM update sequence). The parameters $\lambda_j$ and $\rho_{ji}$ are the maximum allowable step sizes in this scheme.}}
\label{alg:bsdmm}
\begin{algorithmic}[1]
\Procedure{CMF-BSDMM}{$\tA^0, \tS^0$}
\State $\tX_1^0, \tX^0_2 \gets \tA^0, \tS^0$
\For{$j=1,2$}
    \State $\tZ_{ji}^0 \gets \tL_{ji}\tX_j\ \forall i \in \lbrace 1,\dots,M_j\rbrace$
    \State $\tU_{ji}^0 \gets \vec{0}\ \forall i \in \lbrace 1,\dots,M_j\rbrace$
\EndFor
\For{$\mathrm{it} = 0,1,\dots$}
    \For{$j=1,2$}
    	\State $\lambda_j^{\mathrm{it}} \gets j==1? \norm{\tX_2^\mathrm{it} \tX_2^{\mathrm{it}\top}}_s^{-1} :  \norm{\tX_1^{\mathrm{it}+1\top} \tX_1^{\mathrm{it}+1}}_s^{-1}$
	\State $\rho_{ji}^{\mathrm{it}} \gets 2 M_j\, \lambda_j^{\mathrm{it}+1} \norm{\tL_{ji}}_2^2$ 
        \State $\tX_j^{\mathrm{it}+1} \gets\prox_{\lambda_j f,j}\left(\tX_{j}^\mathrm{it}-\sum_{i=1}^{M_{j}}\frac{\lambda_j}{\rho_{ji}}\tL_{ji}^{\top}\left(\tL_{ji}\tX_{j}^\mathrm{it}-\tZ_{ji}^\mathrm{it}+\tU_{ji}^\mathrm{it}\right)\right)$
        \For{$i = 1,\dots,M_j$}
    	    \State $\tZ_{ji}^{\mathrm{it}+1} \gets \prox_{\rho_{ji} g_{ji}}\left(\tL_{ji}\tX_{j}^{\mathrm{it}+1}+\tU_{ji}^\mathrm{it}\right)$
   	    \State $\tU_{ji}^{\mathrm{it}+1} \gets \tU_{ji}^\mathrm{it}+\tL_{ji}\tX_{j}^{\mathrm{it}+1}-\tZ_{ji}^{\mathrm{it}+1}$
	\EndFor
   \EndFor
      \If{$\bigwedge_{ji}\left\lbrace\norm{\tens{P}_{ji}^{\mathrm{it}+1}}_2 \leq \epsilon^\mathrm{pri} \wedge \norm{\tens{D}_{ji}^{\mathrm{it}+1}}_2 \leq \epsilon^\mathrm{dual}\right\rbrace$}
     break
    \EndIf
\EndFor
\EndProcedure
\end{algorithmic}
\end{algorithm}

\subsection{Constraints}
\label{sec:constraints}

The conventional NMF is capable of separating components, but only from the color contrast between objects in the scene.
If the color contrast vanishes or the number $K$ of overlapping components gets large, the model in \autoref{eq:model} becomes increasingly susceptible to degeneracies, where the mixture model yields a fair representation of the scene but many of the individual components look implausible.

We thus prefer the flexibility of the CMF to impose additional priors or constraints based on existing knowledge or intuition about how SEDs and morphologies are distributed.
In addition to non-negativity, which is justified for astronomical sources that emit photons, we will routinely employ symmetry, monotonicity, and sparsity constraints for the galaxy morphologies (see \autoref{sec:sims}).
We currently do not utilize priors on galaxy or stellar SEDs, but we will demonstrate in \autoref{sec:jet} how color constraints can help in cases when additional stability of the source extraction is desired.

We want to emphasize that \scarlet\ is capable of incorporating any prior or constraint that can be expressed as a proximal operator.
We therefore think of it as a general-purpose component separation framework that can be optimized for a wide variety of applications.

\subsubsection{Symmetry}
\label{sec:symmetry}
We take guidance from the SDSS {\it Photo} deblender (Lupton, in prep.), which, inspired by \citet{Doi95.1}, constructs templates of the sources that are symmetric under $180^\circ$ rotations around the peak pixel.
We formalize this requirement by stating that 
\begin{equation}
\forall k,i: \tS_{k,i} - \tS_{k,j} = 0,
\end{equation}
where $i$ and $j$ are the indices of the symmetric partner pixels,%
\footnote{In order to always have a well-defined partner for each pixel (other than the peak pixel), the peak pixel needs to be in the center of the image, and the number of pixels in the image must be odd in both directions.
We achieve this by defining a virtual box around each source that contains all of its pixels, with the peak pixel at the box center.
%In \autoref{sec:centering} we show how the transformation between the virtual source boxes and the original image is done.
}
which can be expressed in matrix form,
\begin{equation}
\forall k: \tL_\mathrm{symm}\tS_k = \vec{0},
\end{equation}
with a symmetry matrix  $\tL_\mathrm{symm}\in \mathbb{R}^{(N/2-1) \times N}$ that has the elements $(1,-1)$ for each of $N/2-1$ suitable pixel pairs (the peak pixel is excluded).
In this linearly transformed space, the constraint is imposed by the projection operator onto the zero vector, i.e. $\prox_0(\tL_\mathrm{symm}\tS_k)$.
The combination of the transformation matrix and the proximal operator is required for any ADMM-style constraint in \autoref{alg:bsdmm}.

Compared to the SDSS {\it Photo} deblender, which adopts the minimum value for both pixels in the symmetric pair, this operator will converge to the mean of both pixels.
This is a direct consequence of the proximal operator being the Euclidean projection operator and the mean statistic having the least squared error.

\subsubsection{Monotonicity}
\label{sec:monotonicity}
The prominent failure mode of the SDSS {\it Photo} deblender arises from its templates not necessarily declining monotonically from the peak pixel.%
\footnote{To address this weakness, the implementation of the SDSS {\it Photo} deblender in the current HSC software stack imposes a monotonicity constraint on previously smoothed templates.}
While resolved images of real galaxies may in detail show non-monotonic behavior, for instance due to star-forming regions in the outer parts of spiral galaxies, we prefer imposing this constraint for the bulk source shape.
An area that violates this constraint likely constitutes a different stellar population, potentially with a color different from the bulk model, and should thus be fit by adding another (monotonic) component to the mixture model.

Monotonicity for two-dimensional features is not uniquely defined.
We use a form of radial monotonicity, i.e.
\begin{equation}
\forall k,i: \sum_{j\in\mathcal{U}(i)} w_{ji} \tS_{k,j} - \tS_{k,i} \geq 0,
\end{equation}  
where $\mathcal{U}(i)$ is the set of neighbor pixels of $i$ that are closer to the peak than $i$, and the weights are normalized such that $\forall i: \sum_j w_{ji}=1$.
We implement two variants:
\begin{enumerate}[label=\arabic*., ref=\arabic*., leftmargin=*, partopsep=0em, itemsep=0em ]
\item The neighborhood contains only the nearest neighbor of $i$ in the direction of the peak, $\mathcal{U}(i) = \lbrace \mathrm{NN}(i) \rbrace$ and $w_{\mathrm{NN}(i),i} = 1$.
\item The neighborhood contains all three neighbor pixels that are closer to the peak, and their weights are given by $w_{ji} = \cos\left(\phi_j - \phi_i\right)$, where $\phi_i$ denotes the angle from the center of pixel $i$ to the center of the peak pixel.
\end{enumerate}
Unsurprisingly, we find the latter variant to yield models that are somewhat softened in azimuthal direction, and they can violate monotonicity along the straight path from pixel $i$ to the peak, which variant 1 would obey.

As before, we rewrite the constraint in matrix form, $\forall k: \tL_\mathrm{mono} \tS_k \geq 0$, which can be enforced by the non-negativity constraint from \autoref{eq:prox_plus_unity} in the transformed domain, $\prox_+(\tL_\mathrm{mono} \tS)$.

This enforcement of monotonicity can be slow to converge in heavily blended regions.
The \scarlet\  package therefore contains a projection operator in the direct domain, $\prox_\mathrm{mono}$,  which constructs a monotonic version of its argument. While this construction will in general not yield the  monotonic solution that is closest to the argument (the operator is thus in fact not a true proximal operator), we have found it to be substantially more robust and efficient in practice.

\subsubsection{Sparsity}
\label{sec:sparsity}

Sparsity refers to a solution that can be represented with a small number or a low amplitude of coefficients in a particular domain.
In the direct, i.e. the pixel, domain, sparsity constraints thus induce solutions with few non-zero pixels and/or compressed pixel amplitudes, which can be useful to reduce the impact of pixel noise.
They are introduced by minimizing the objective function and a penalty function based on the norm of the solution, often the $\ell_2$ or $\ell_1$ norms or the $\ell_0$ pseudo-norm, or a combination of them.
While the $\ell_2$ norm is differentiable, the other two are not. However, both $\ell_0$ and $\ell_1$ have simple proximal operators, known as ``hard thresholding'' and ``soft thresholding'' operators, respectively:
\begin{equation}
\begin{split}
&\prox_{\lambda \ell_0}(\x) = (x_1^h, \dots, x_n^h)\ \text{with}\ x_i^h = \begin{cases}0 & \text{if}\ |x_i|<\lambda \\ x_i & \text{else}\end{cases}\\
&\prox_{\lambda \ell_1}(\x) =  (x_1^s, \dots, x_n^s)\ \text{with}\ x_i^s = \text{sign}(x_i)\,\text{max}\left(0, |x_i|-\lambda\right)
\end{split}
\end{equation}
This is the first example of a proximal operator of a function other than an indicator of a closed set.
The parameter $\lambda$ then governs the strength of the penalty compared to the objective function $f$.

Because \scarlet\ allows arbitrary transformations of $\tA$ and $\tS$ with linear operators $\tL$, the sparsity constraints can be applied in the direct domain and/or in the transformed domain. 

\subsubsection{Other options}

We do not attempt to provide a comprehensive list of possible constraints that are expressible by proximal operators. Instead, we discuss several that we found useful and refer the reader to the appendix of \citet{Combettes2009} and section 6 of \citet{Parikh2014} for more details.

Spatially flat components, e.g. for fitting a uniform sky emission, can be realized by projecting onto the mean (see discussion in \autoref{sec:symmetry}): $\prox_\mathrm{flat}(\x) = \tfrac{1}{n}\sum_i^n x_i \vec{1}_n$ with $\x\in\mathbb{R}^n$. 

To allow some amount of spatial variability, e.g. for describing Intra-Cluster Light, we find the Maximum Entropy regularization \citep{Frieden1972} useful, which fundamentally seeks to find the flattest distribution that is compatible with the data by adding the (negative) Shannon entropy term
\begin{equation}
S = \sum_i^N p_i \log(p_i)\ \text{with}\ p_i = x_i \Big/ \sum x_i
\end{equation}
to the objective function $f$.
The proximal operator of $S$ is 
\begin{equation}
\text{prox}_{\lambda S}(\x) = \lambda \Re\left[W\left(\frac{1}{\lambda} \exp\left(\frac{\x}{\lambda}-1\right)\right)\right],
\end{equation}
where $W$ denotes the Lambert-$W$ function.

%Another option is the Total Variation \citep[TV,][]{Rudin1992} regularization, which seeks to create large homogeneous areas while preserving their edges.
%The penalty term is of the form $V(\x) = \sum_i \norm{x_{i+1} - x_i}_1$, which can easily be expressed by a first-order forward difference matrix $\tL_{\Delta x}$, passed to $\prox_{\ell_1}(\tL_{\Delta x}\x)$.
%In other words, the solution shall have a sparse gradient.
%For two-dimensional images, the ``anisotropic TV'' penalty can be split into two terms for the horizontal and the vertical gradients, which best preserves edges that are aligned with the coordinate axes.

By choosing the matrix $\tL$ to represent a basis transform, one can fundamentally alter the characteristics of the solution one wants to isolate or penalize.
A few examples with applications in data analysis are the Fourier transform \citep{Hassanieh2012}, the wavelet transform \citep{Rapin2014}, or the shapelet transform \citep{Refregier2003}.
In the transformed domain, simple projection operators can then e.g. null undesired frequencies.

Finally, instead of hard constraints one can also introduce prior distributions on the solutions.
If the prior distribution $p(\x)$ is smooth, the corresponding proximal operator is given by the gradient step of $\log(p)$ analogous to \autoref{eq:prox_f}; if it has hard edges it can be represented by wrapping the gradient step in a proximal gradient update as in \autoref{eq:pgm}.

\subsection{Heteroscedastic and correlated errors}
\label{sec:weights}

\autoref{eq:simple_f} does not account for heteroscedastic errors.
\citet{Blanton2007,Zhu2016} introduced a simple extensions to heteroscedastic errors with the matrix $\tens{V}$ of variances of each pixel in $\tY$:
\begin{equation}
\label{eq:weighted_f}
f(\tens{A},\tens{S})=\norm{V^{-\frac{1}{2}}\circ(\tens{A}\cdot\tens{S} - \tens{D})}_2^2,
\end{equation}
where $\circ$ denotes the element-wise, or Hadamard, product.
It is straightforward to adjust the gradients in \autoref{eq:gradients} but it is not clear how their Lipschitz constants differ from \autoref{eq:lipschitz}.%
\footnote{The two works cited above did not need to compute the respective Lipschitz constants as they solved the NMF by multiplicative updates, not proximal algorithms.}
In addition, \autoref{eq:weighted_f} does not account for correlations in the noise that arise when images are astrometrically rectified.

For that purpose it is necessary to serialize $\tY$, $\tA$, and $\tS$ into vectors ($\vec{Y},\vec{A},\vec{S}$) with lengths $B\cdot N$, $B\cdot K$ and $K\cdot N$, respectively.
The most convenient vector form $\vec{Y} = \left(\tY_{1}, \dots, \tY_{B}\right)$ simply concatenates the images in $B$ bands.
The covariance matrix of $\vec{Y}$ has block-diagonal form
\begin{equation}
\tens{\Sigma}_\vec{Y} = \begin{pmatrix}
\tens{\Sigma}_1 & \tens{0} & \dots & \tens{0}\\
\tens{0} & \tens{\Sigma}_2 & \dots & \tens{0}\\
\vdots & \vdots & \ddots & \tens{0}\\
\tens{0} & \tens{0} & \dots & \tens{\Sigma}_B\\
\end{pmatrix},
\end{equation}
with $\tens{\Sigma}_b\in \mathbb{R}^{N\times N}$ denoting the pixel covariance matrix in band $b$.
We can now rewrite the objective function in a form that uses vectors for $\tY$ and one of $\tA$ or $\tS$, with the other one expressed by a sparse block matrix ($\bar{\tA}$ or $\bar{\tS}$).
The objective function is then
\begin{equation}
\label{eq:covariance_f}
\begin{split}
&f(\vec{A},\tS) = \frac{1}{2}\norm{(\vec{Y}-\bar{\tS}\vec{A})^\top\tens{\Sigma}^{-1}_\vec{Y} (\vec{Y}-\bar{\tS}\vec{A})}_2^2,\\
&f(\tA,\vec{S}) = \frac{1}{2}\norm{(\vec{Y}-\bar{\tA}\vec{S})^\top\tens{\Sigma}^{-1}_\vec{Y} (\vec{Y}-\bar{\tA}\vec{S})}_2^2\\
\end{split}
\end{equation}
which describes an ordinary vector-valued linear inverse problem.
The gradients are
\begin{equation}
\begin{split}
\nabla_\vec{A} f(\vec{A},\tS) &= -\bar{\tS}^\top\tens{\Sigma}_\vec{Y}^{-1}(\vec{Y}-\bar{\tS}\vec{A})\\
\nabla_\vec{S} f(\tA,\vec{S}) &= -\bar{\tA}^\top\tens{\Sigma}_\vec{Y}^{-1}(\vec{Y}-\bar{\tA}\vec{S})\\
\end{split}
\end{equation}
and the covariance matrices are given by
\begin{equation}
\begin{split}
\tens{\Sigma}_\vec{A} &= \bar{\tS}^\top\tens{\Sigma}_\vec{Y}^{-1}\bar{\tS}\\
\tens{\Sigma}_\vec{S} &= \bar{\tA}^\top\tens{\Sigma}_\vec{Y}^{-1}\bar{\tA}.
\end{split}
\end{equation}
The Lipschitz constants of those gradients, which we need to calculate $\lambda_j$ in \autoref{alg:bsdmm}, are the spectral norms of the covariance matrices.
Finally, the block matrices that encode $\tA$ or $\tS$ are given by
\begin{equation}
\bar{\tA} = \begin{pmatrix}
\tA_{11} \tens{I}_N & \tA_{21} \tens{I}_N & \dots & \tA_{K1} \tens{I}_N\\
\tA_{12} \tens{I}_N & \tA_{22} \tens{I}_N & \dots & \tA_{K2} \tens{I}_N\\
\vdots & \vdots & \ddots & \vdots\\
\tA_{1B} \tens{I}_N & \tA_{2B} \tens{I}_N & \dots & \tA_{KB} \tens{I}_N\\
\end{pmatrix} \in \mathbb{R}^{BN\times KN}
\end{equation}
and 
\begin{equation}
\label{eq:Sbar}
\bar{\tS} = \begin{pmatrix}
\tS^\top & \tens{0} & \dots & \tens{0}\\
\tens{0} & \tS^\top & \dots & \tens{0}\\
\vdots & \vdots & \ddots & \tens{0}\\
\tens{0} & \tens{0} & \dots & \tS^\top\\
\end{pmatrix} \in \mathbb{R}^{BN\times BK},
\end{equation}
each stacking the blocks row-wise over multiple bands.

\subsection{PSF (de-)convolution}
\label{sec:psf}

Incorporating the convolution with the point spread function (PSF) into our model is particularly useful for a deblender because it allows a distinction between overlap that is caused by the physical extent of the sources involved and the one due to atmospheric or instrumental blurring of the images.
Knowing the latter reduces the ambiguity caused by the former.

In addition, our model from \autoref{eq:model} fits the same morphology to each band, which
is only meaningful if the PSF does not change between bands or the images have been PSF-homogenized across all bands.
Otherwise, the data will exhibit color variations that our model cannot reproduce, even if the underlying morphological structure of the astronomical source is achromatic like e.g. that of a star.

In the vectorized frame we have introduced in \autoref{sec:weights} we express the convolution as
\begin{equation}
\bar{\tens{P}} = \begin{pmatrix}
\tens{P}_1 & \tens{0} & \dots & \tens{0}\\
\tens{0} & \tens{P}_2 & \dots & \tens{0}\\
\vdots & \vdots & \ddots & \tens{0}\\
\tens{0} & \tens{0} & \dots & \tens{P}_B\\
\end{pmatrix} \in \mathbb{R}^{BN\times BN},
\end{equation}
with each of the blocks describing the PSF in a given band as a linear operator that couples pixels of $\vec{S}$.
The model can thus be extended to have the objective functions
\begin{equation}
\label{eq:convolution_f}
\begin{split}
&f(\vec{A},\tS) = \frac{1}{2}\norm{(\vec{Y}-\bar{\tens{P}}\bar{\tS}\vec{A})^\top\tens{\Sigma}^{-1}_\vec{Y} (\vec{Y}-\bar{\tens{P}}\bar{\tS}\vec{A})}_2^2,\\
&f(\tA,\vec{S}) = \frac{1}{2}\norm{(\vec{Y}-\bar{\tens{P}}\bar{\tA}\vec{S})^\top\tens{\Sigma}^{-1}_\vec{Y} (\vec{Y}-\bar{\tens{P}}\bar{\tA}\vec{S})}_2^2\\
\end{split}
\end{equation}
with gradients
\begin{equation}
\label{eq:full_gradients}
\begin{split}
\nabla_\vec{A} f(\vec{A},\tS) &= -\left(\bar{\tens{P}}\bar{\tS}\right)^\top\tens{\Sigma}_\vec{Y}^{-1}(\vec{Y}-\bar{\tens{P}}\bar{\tS}\vec{A})\\
\nabla_\vec{S} f(\tA,\vec{S}) &= -\left(\bar{\tens{P}}\bar{\tA}\right)^\top\tens{\Sigma}_\vec{Y}^{-1}(\vec{Y}-\bar{\tens{P}}\bar{\tA}\vec{S})\\
\end{split}
\end{equation}
and covariance matrices
\begin{equation}
\begin{split}
\tens{\Sigma}_\vec{A} &= \left(\bar{\tens{P}}\bar{\tS}\right)^\top\tens{\Sigma}_\vec{Y}^{-1}\bar{\tens{P}}\bar{\tS}\\
\tens{\Sigma}_\vec{S} &= \left(\bar{\tens{P}}\bar{\tA}\right)^\top\tens{\Sigma}_\vec{Y}^{-1}\bar{\tens{P}}\bar{\tA},
\end{split}
\end{equation}
from which we can compute $\prox_f$ and the step sizes $\lambda$.

This ansatz with the PSF as a linear operator is only valid as long as the PSF images and the deconvolved model remain well sampled.%
\footnote{Internal oversampling is an option to relax this requirement, which we will investigate in future work.}
Otherwise, the discretized nature of either will lead to convolution artifacts, which may result, e.g., in apparent non-monotonic behavior in the deconvolved morphology even if the source is perfectly monotonic.

We therefore recommend to construct difference kernels for each band, $\tens{P}_{\Delta,b} \tens{P}_\mathrm{min} = \tens{P}_b$, such as to perform a partial deconvolution, resulting in an image with a band-independent PSF $\tens{P}_\mathrm{min}$.
Requiring Nyquist-sampling for $\tens{P}_{\Delta,b}$ and $\tens{P}_\mathrm{min}$ suggest that $\tens{P}_\mathrm{min}$ should have a width $T$ of about $\tfrac{1}{2} \min_b T(\tens{P}_b)$.

\subsection{Translations and centering}
\label{sec:centering}

In blends, peak positions from the initial detection step may not accurately reflect the true source centers.
In fact, the apparent peak positions are often shifted towards neighboring objects.
In addition, the combination of symmetry and monotonicity constraints results in the model being strongly affected by the assumed location of each source's center.
To avoid spurious dipole residuals from inaccurate center positions, we introduce linear translation operators $\tens{T}_x$ and $\tens{T}_y$ $\in \mathbb{R}^{N\times N}$ to shift the position of each source.
These operators encode the translation operation, e.g.
\begin{equation}
\forall i<N: (\tS_{k,i}, \tens{S}_{k,i+1}) \rightarrow ((1-dx)\,\tens{S}_{k,i}, dx\,\tens{S}_{k,i+1})
\end{equation}
for a shift of component $k$ by an amount $dx$ in the positive horizontal direction, as sparse band-diagonal matrices, i.e. we treat the translation as a linear interpolation between the pixel and its relevant neighbors.
This is again feasible only if the source is well sampled, and higher-order interpolation schemes can be implemented if needed.

We introduce the translation in horizontal and vertical directions into the model via $\mathbf{S}^\prime = \bar{\tens{T}} \mathbf{S}$, using the block-diagonal matrix
\begin{equation}
\bar{\tens{T}} = \begin{pmatrix}
\tens{T}_1 & \tens{0} & \dots & \tens{0}\\
\tens{0} & \tens{T}_2 & \dots & \tens{0}\\
\vdots & \vdots & \ddots & \tens{0}\\
\tens{0} & \tens{0} & \dots & \tens{T}_K\\
\end{pmatrix} \in \mathbb{R}^{KN\times KN},
\end{equation}
where the two-dimensional shift for each component is given by $\tens{T}_k = \tens{T}_{y,k} \tens{T}_{x,k}$.
If multiple components belong to the same source, they have identical $\tens{T}_k$.
The shifted morphology vector $\mathbf{S}^\prime$ is then used instead of $\mathbf{S}$ in \autoref{eq:covariance_f}, \autoref{eq:Sbar}, and \autoref{eq:convolution_f}.
This way, all gradients and Lipschitz constants are correctly computed with the equations given above.

To determine optimal source centers, we create a new set of $\tens{T}^\prime_{x,k}$ and $\tens{T}^\prime_{y,k}$ matrices, each with an additional shift $\delta$ of typically $0.1\,$pix in either horizontal or vertical direction.
We form two difference images for each component,
\begin{equation}
\begin{split}
\tens{D}_{x,k} &= \tens{A}_k  \times \left( \tens{T}^\prime_{x,k} - \tens{T}_{k} \right) \tens{S}_k\\
\tens{D}_{y,k} &= \tens{A}_k  \times \left( \tens{T}^\prime_{y,k} - \tens{T}_{k} \right) \tens{S}_k,\\
\end{split}
\end{equation}
and combine them into the matrix
\begin{equation}
    \bar{\tens{D}} =
    \begin{pmatrix}
        \tens{D}_{x,1} & \tens{D}_{y,1} & \tens{D}_{x,2} & \tens{D}_{y,2} & \dots & \tens{D}_{x,K} & \tens{D}_{y,K}
    \end{pmatrix}.
\end{equation}
We then state that the current model residuals are a linear combination of the difference images,
\begin{equation}
     \vec{Y}-\bar{\tens{P}}\bar{\tA}\vec{S} = \bar{\tens{D}} \mathbf{\Delta},
\end{equation}
an ordinary linear inverse problem for the two-dimensional offsets $\mathbf{\Delta}$, which we solve with a least-squares solver.
We perform this secondary minimization every 10-th iteration, so that the updates for $\tS$ have time to adjust to changes in the centers.

\section{Applications}
\label{sec:results}

\subsection{Scene from the HSC UltraDeep survey}
\label{sec:hsc}

\begin{figure*}[ph]
    \includegraphics[width=.247\linewidth]{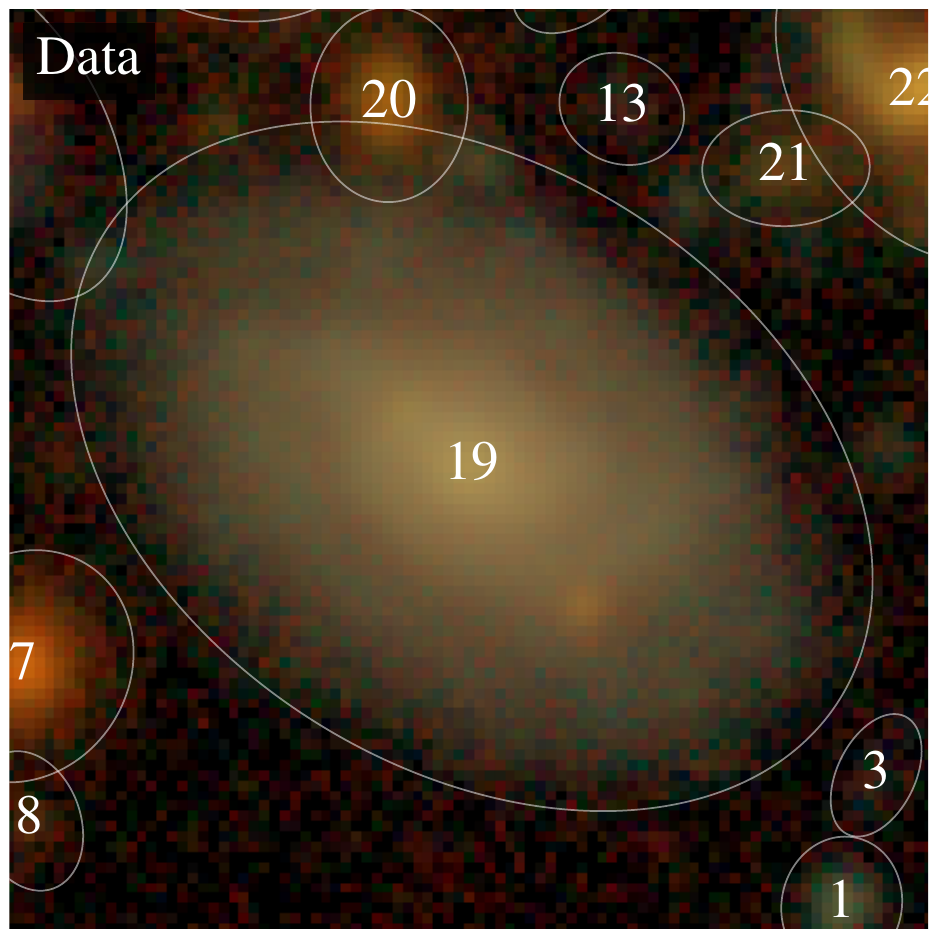}
    \includegraphics[width=.247\linewidth]{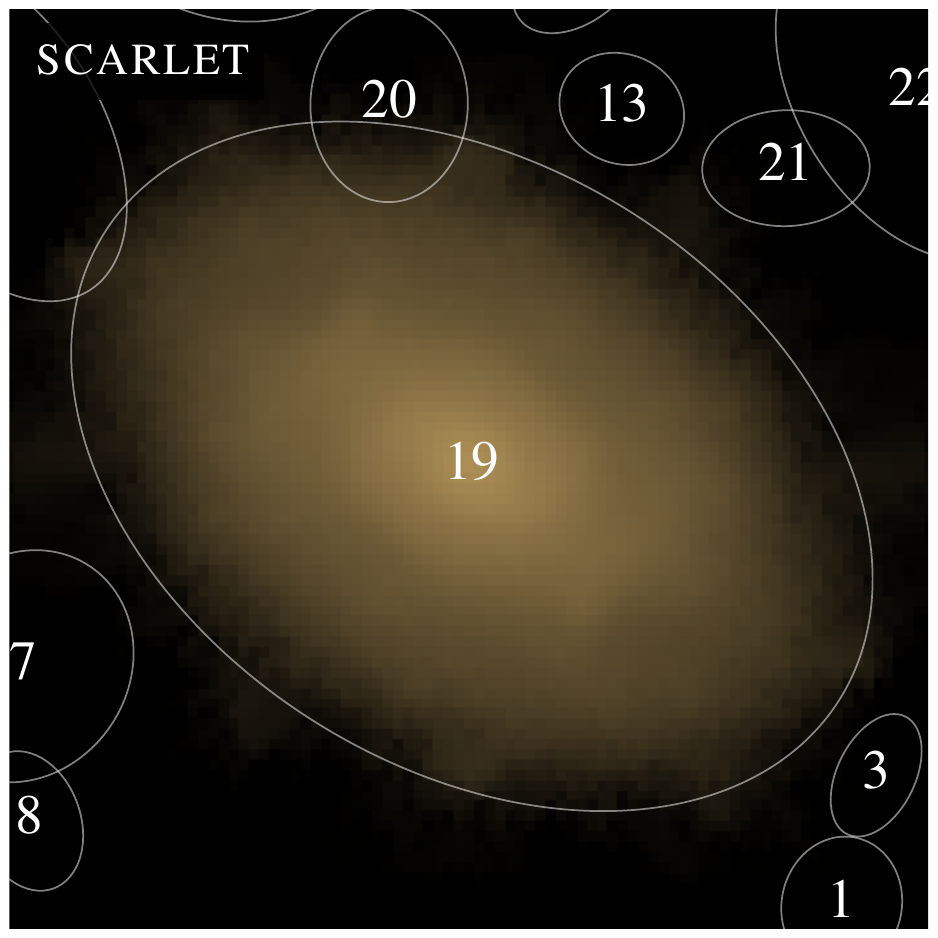}
    \includegraphics[width=.247\linewidth]{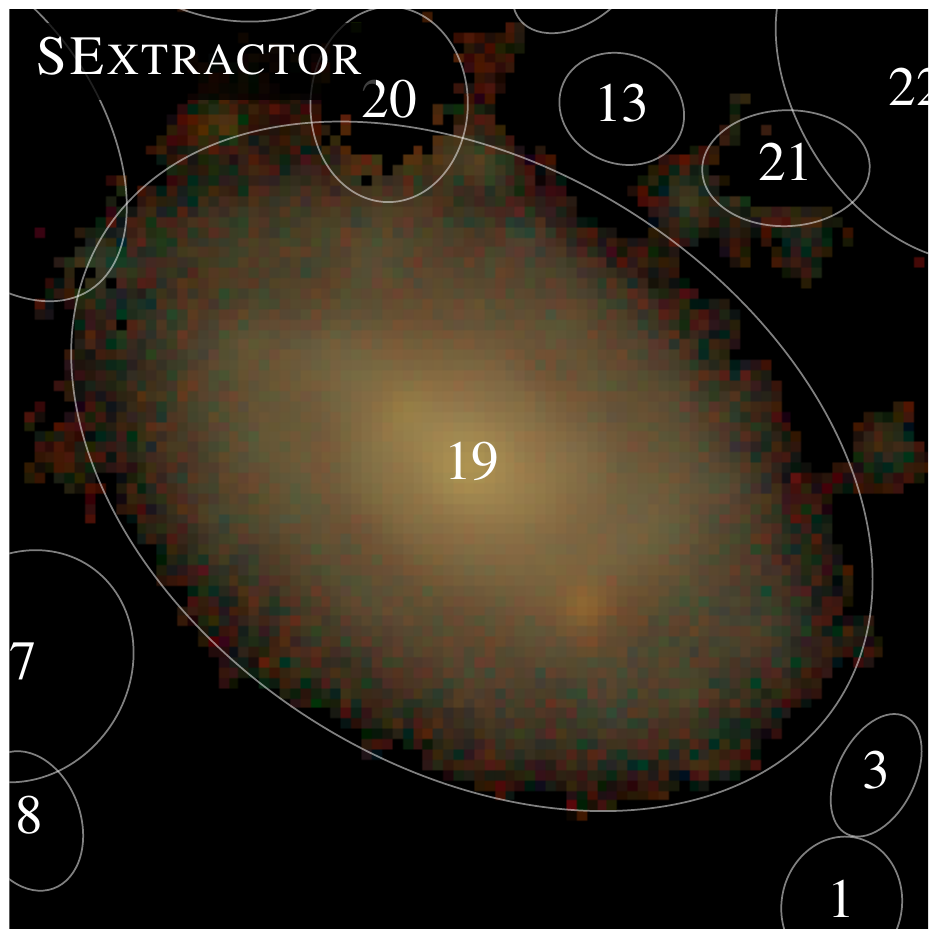}
    \includegraphics[width=.247\linewidth]{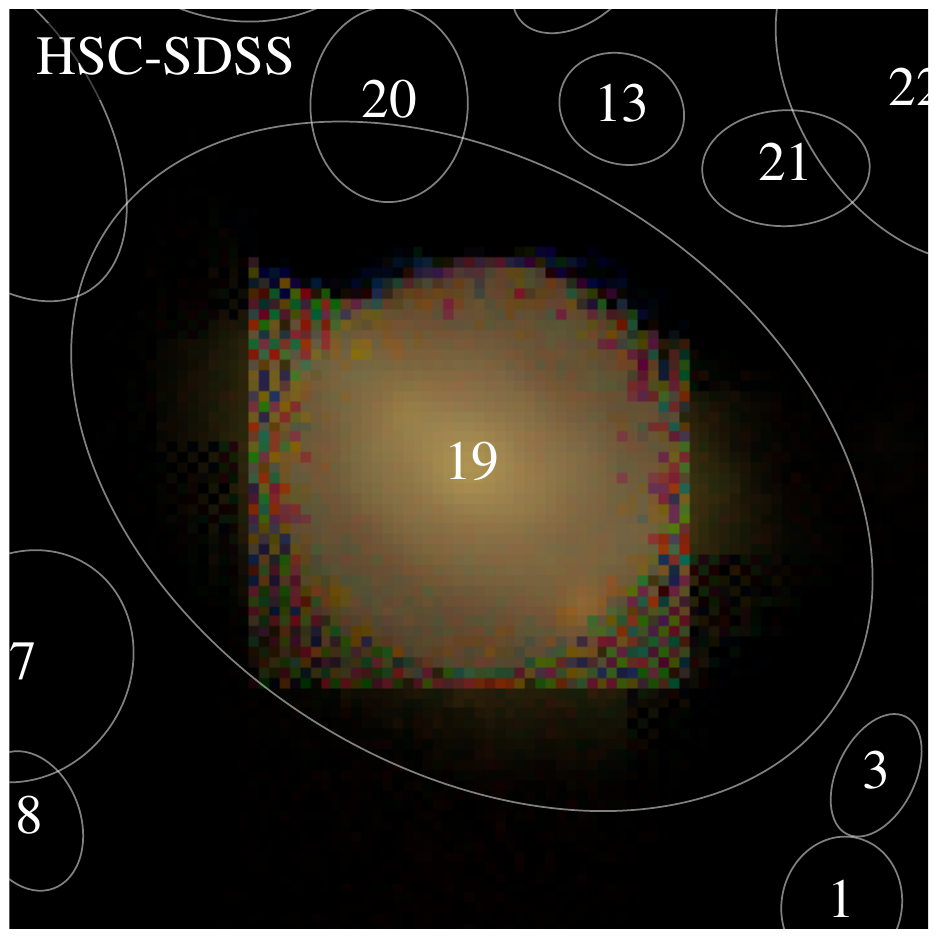}\\
    \includegraphics[width=.247\linewidth]{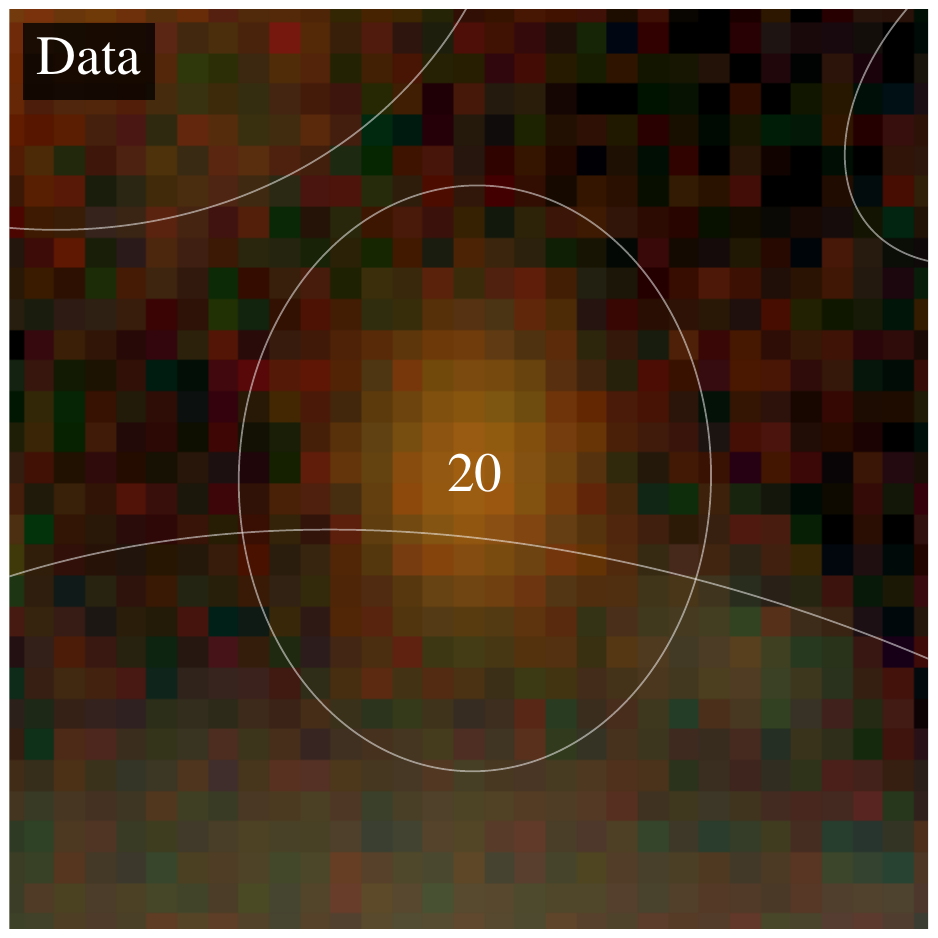}
    \includegraphics[width=.247\linewidth]{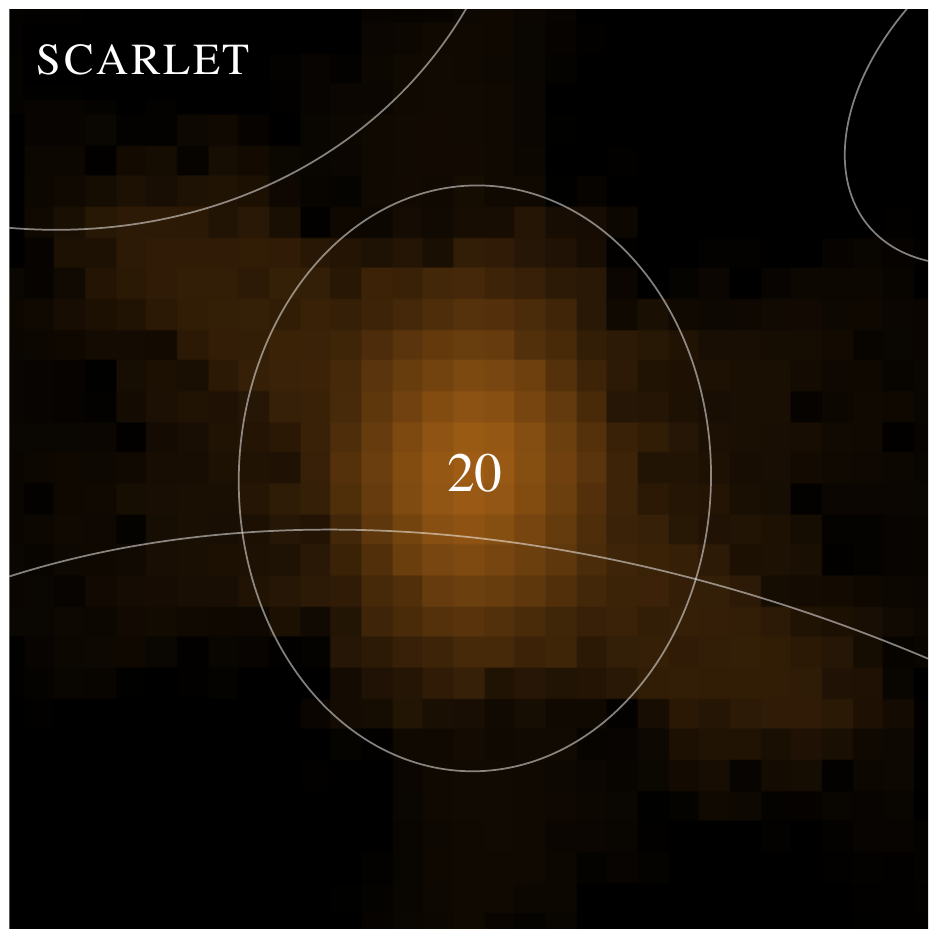}
    \includegraphics[width=.247\linewidth]{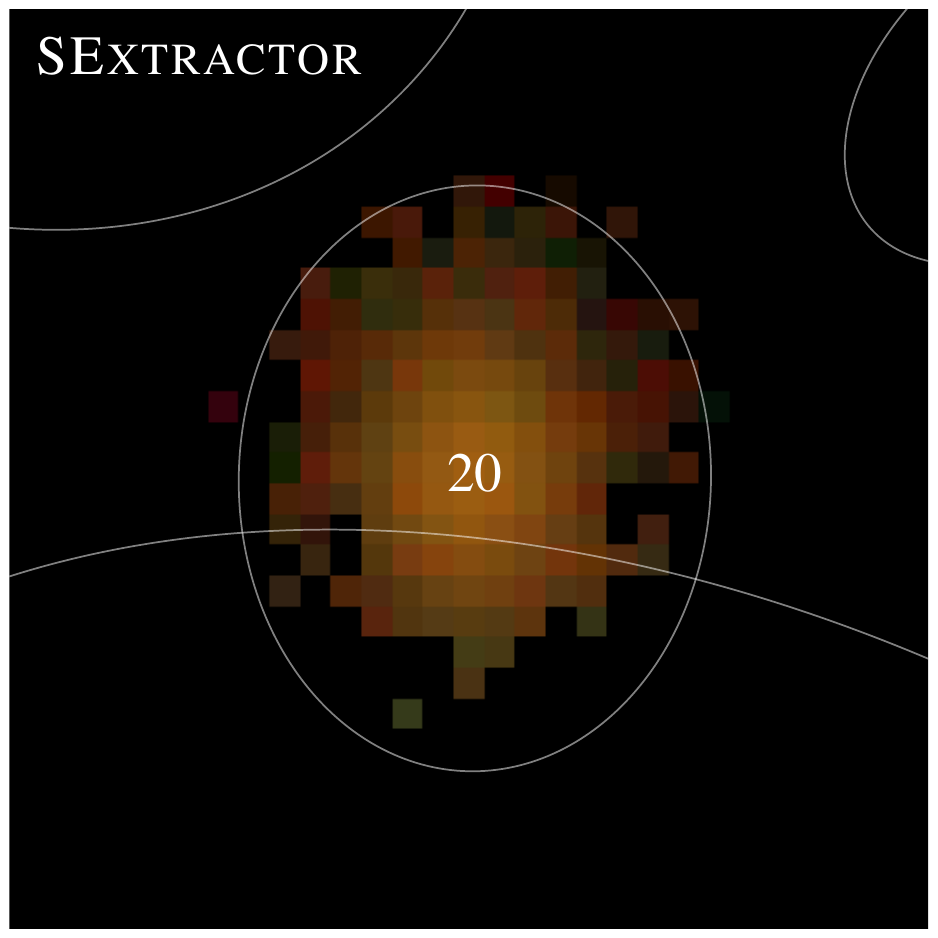}
    \includegraphics[width=.247\linewidth]{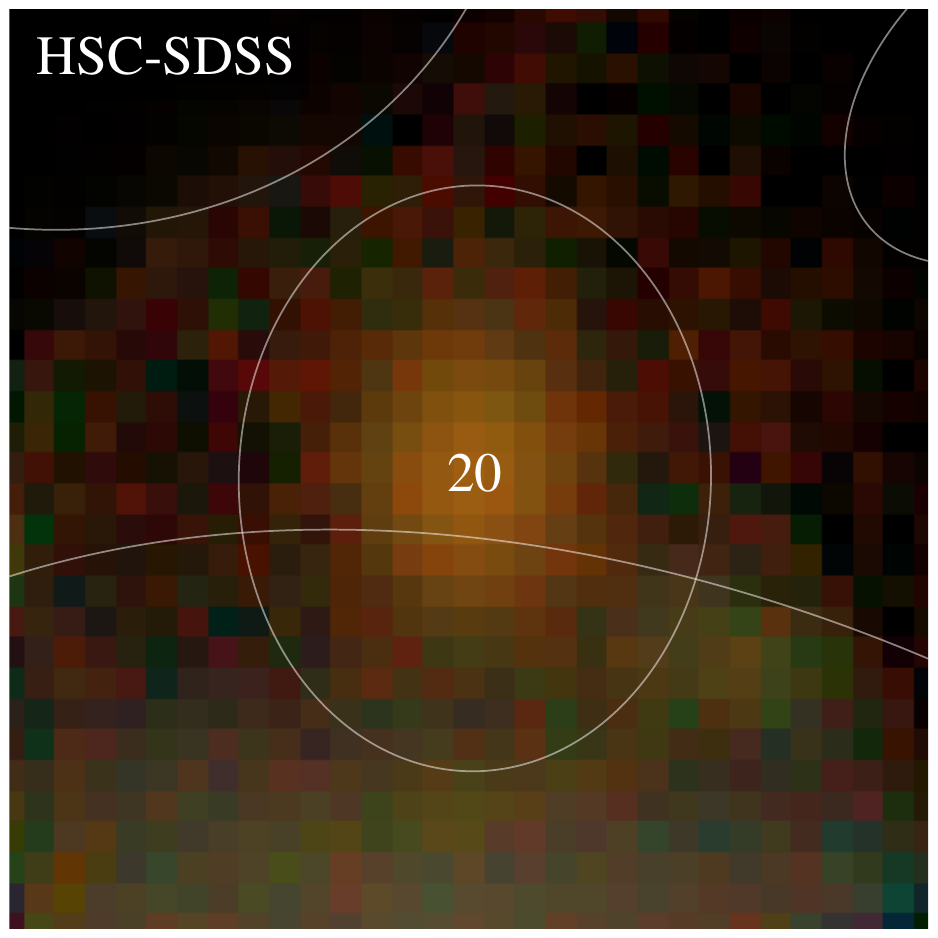}\\
    \includegraphics[width=.247\linewidth]{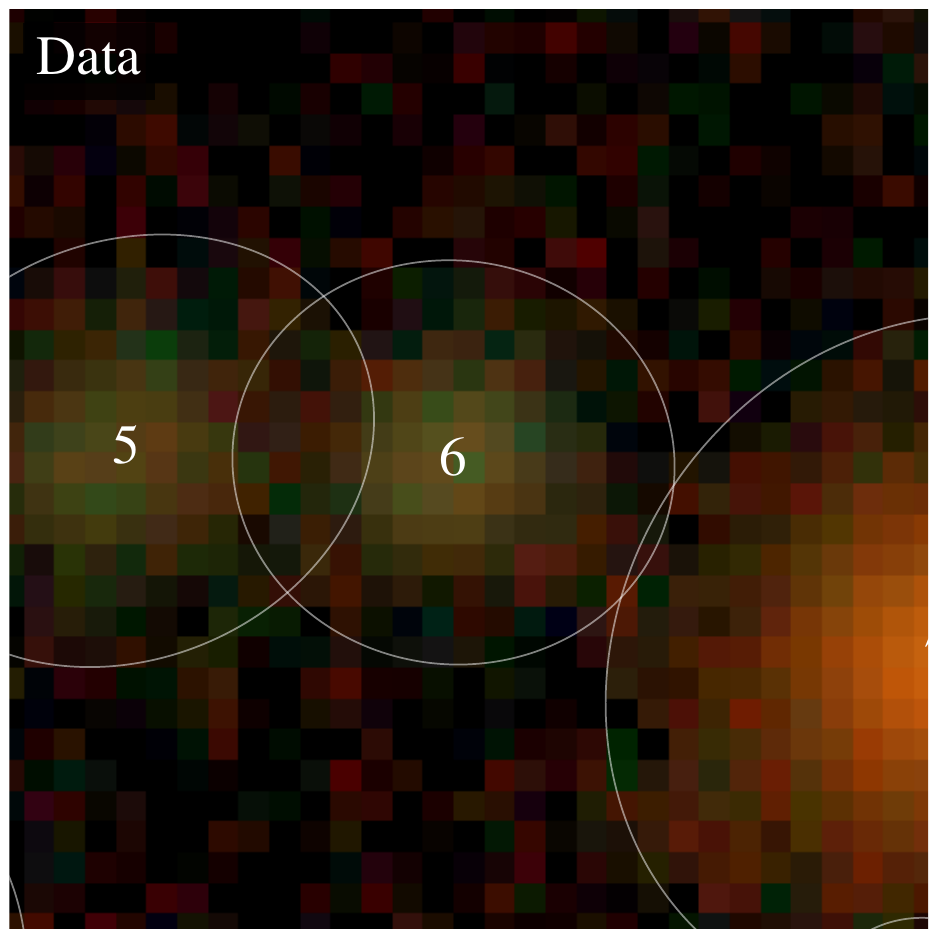}
    \includegraphics[width=.247\linewidth]{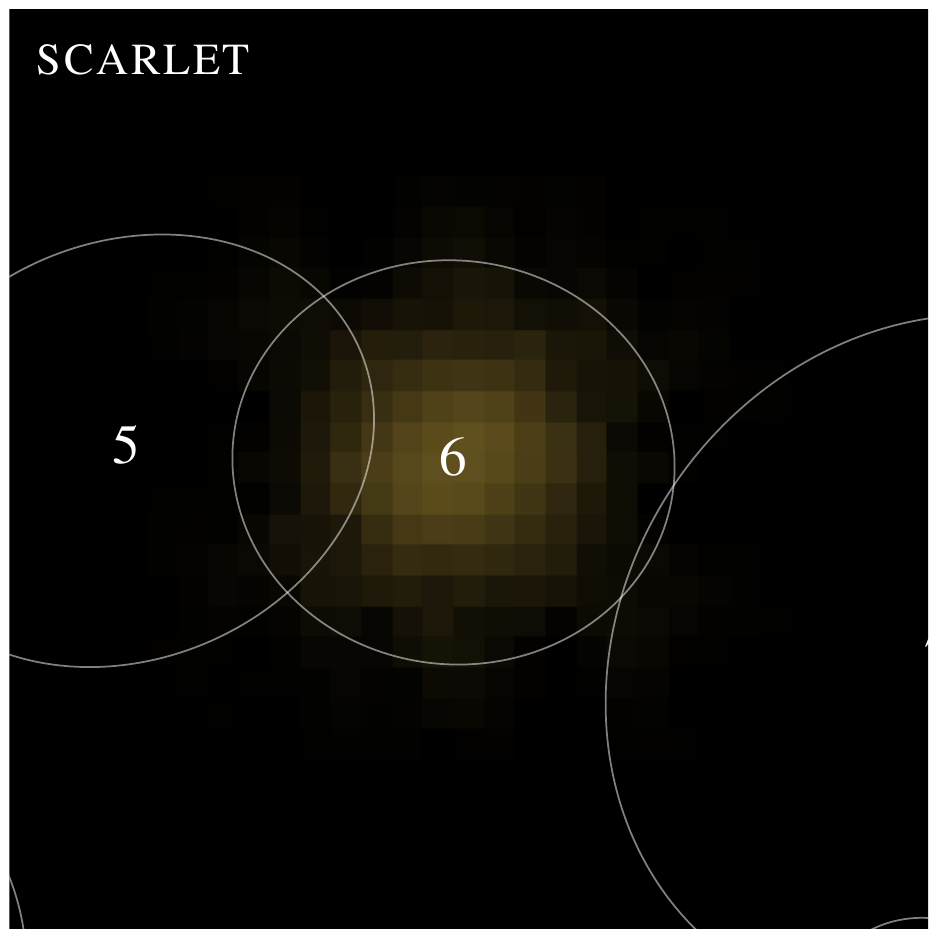}
    \includegraphics[width=.247\linewidth]{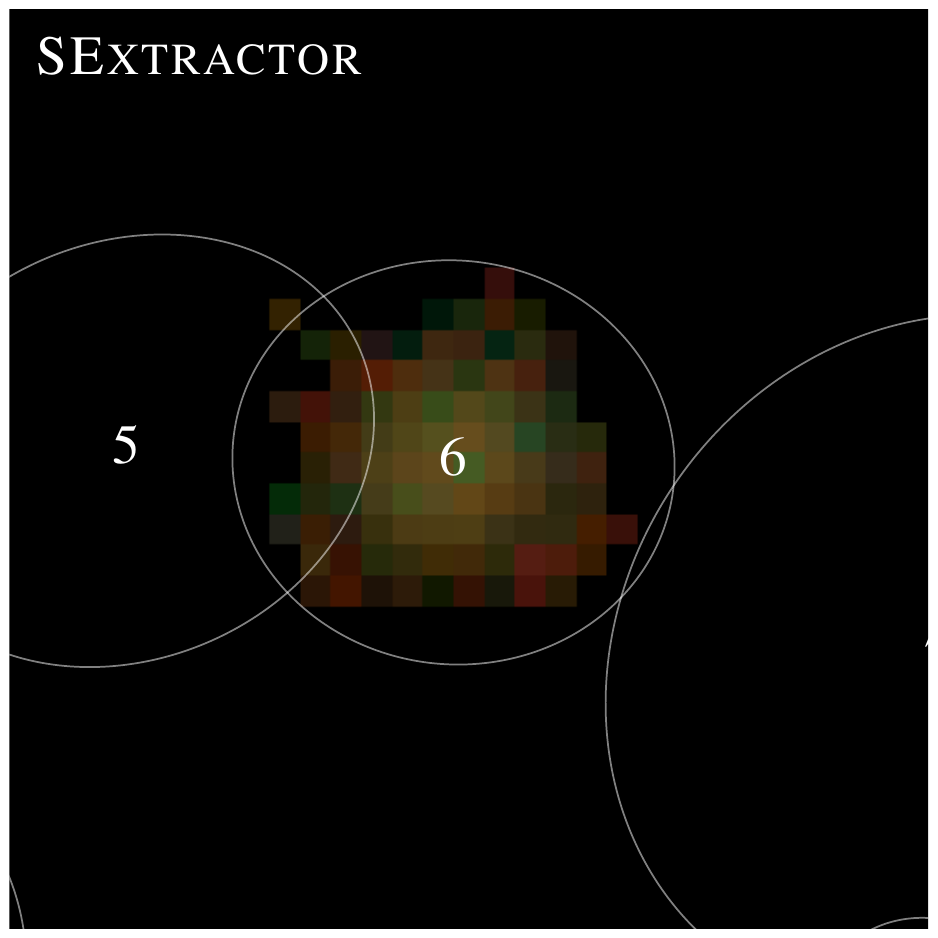}
    \includegraphics[width=.247\linewidth]{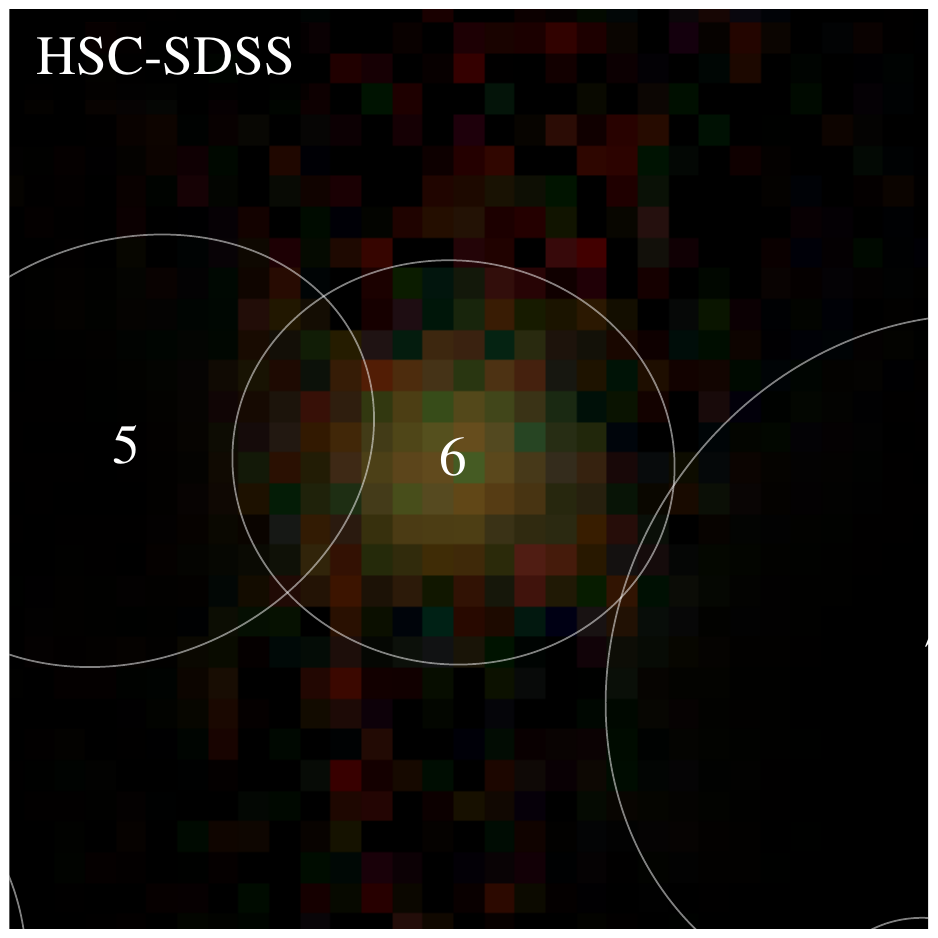}\\
    \includegraphics[width=.247\linewidth]{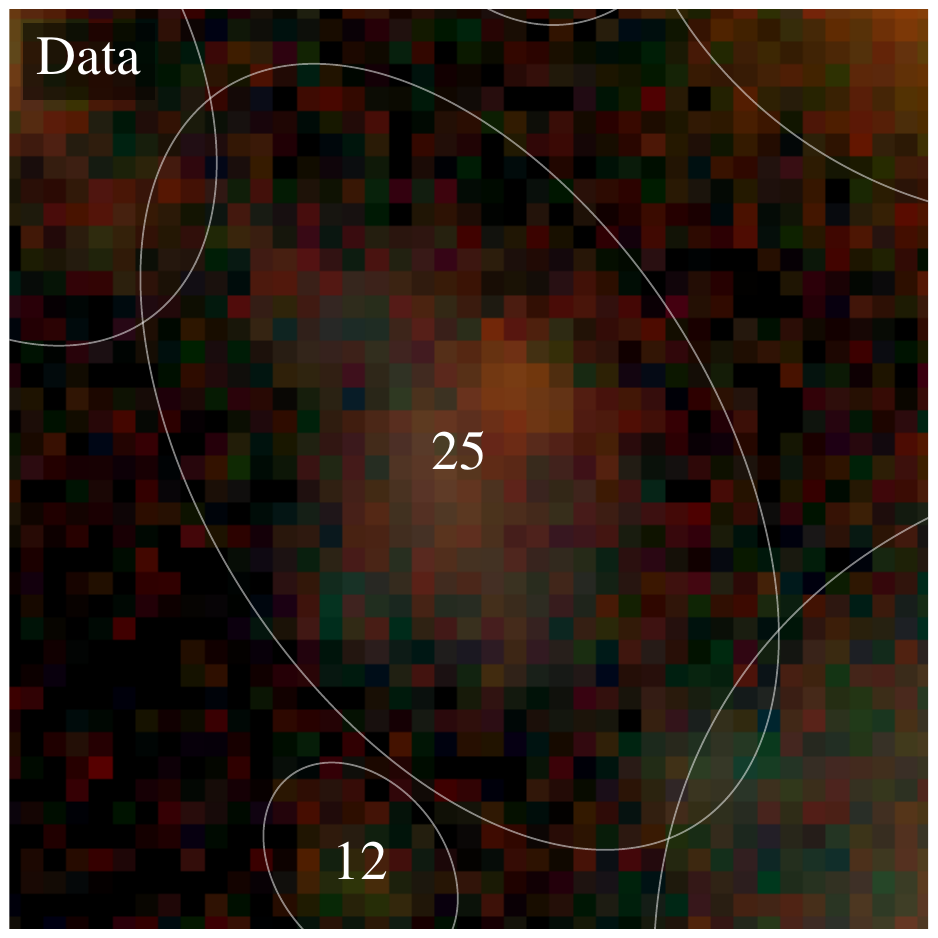}
    \includegraphics[width=.247\linewidth]{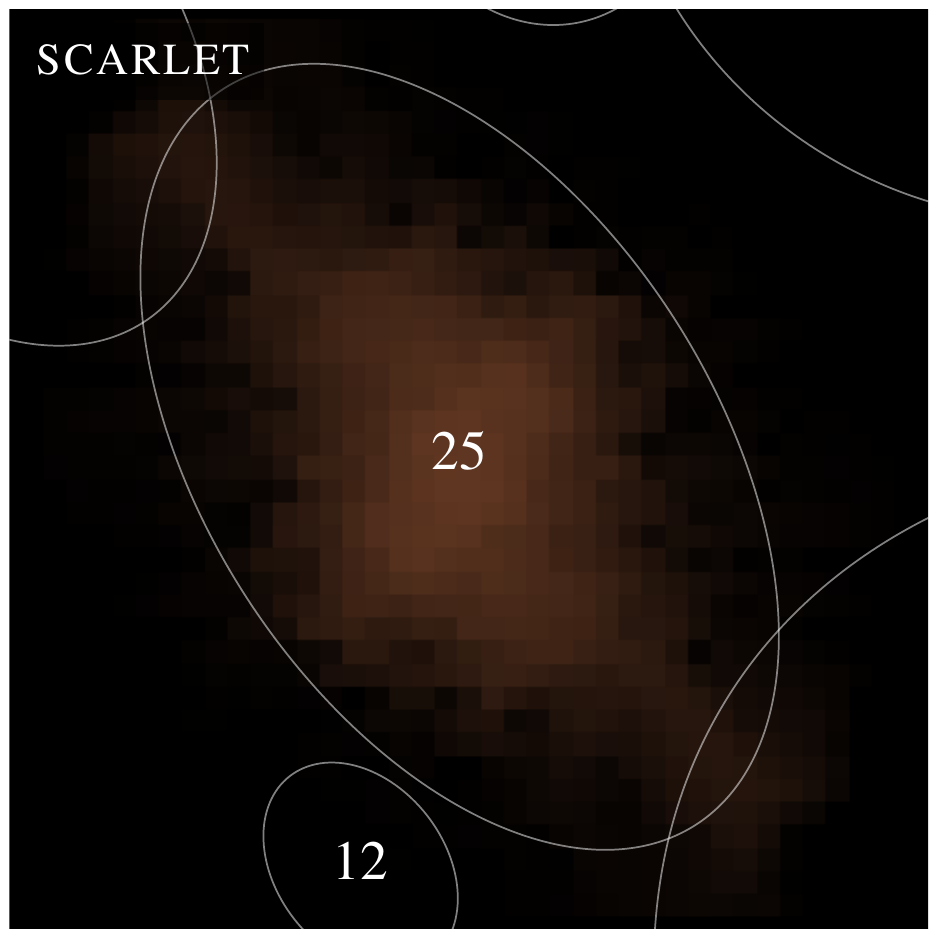}
    \includegraphics[width=.247\linewidth]{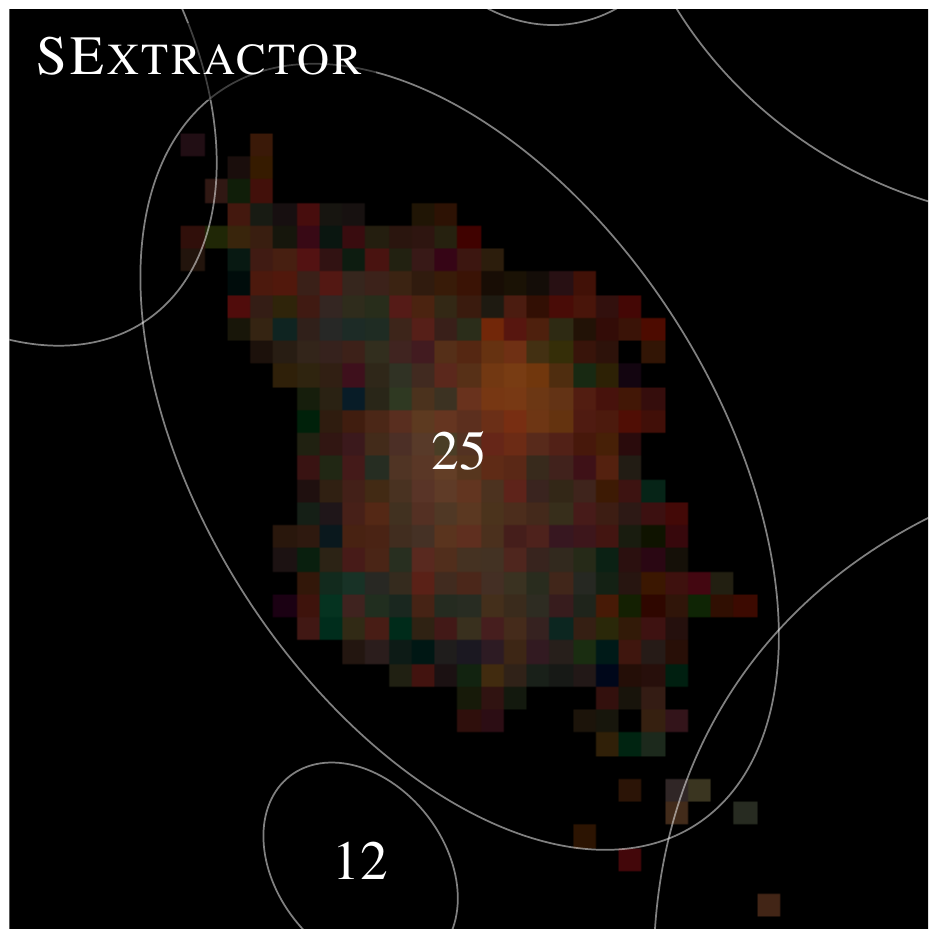}
    \includegraphics[width=.247\linewidth]{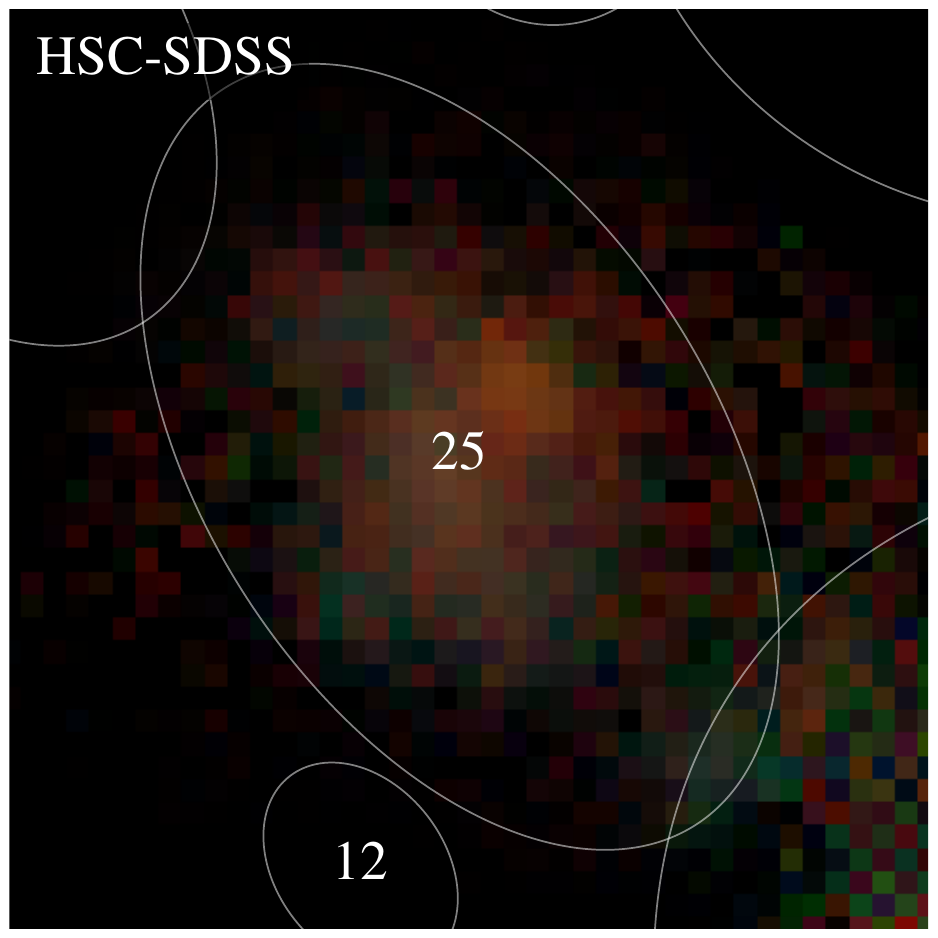}\\
    \includegraphics[width=.247\linewidth]{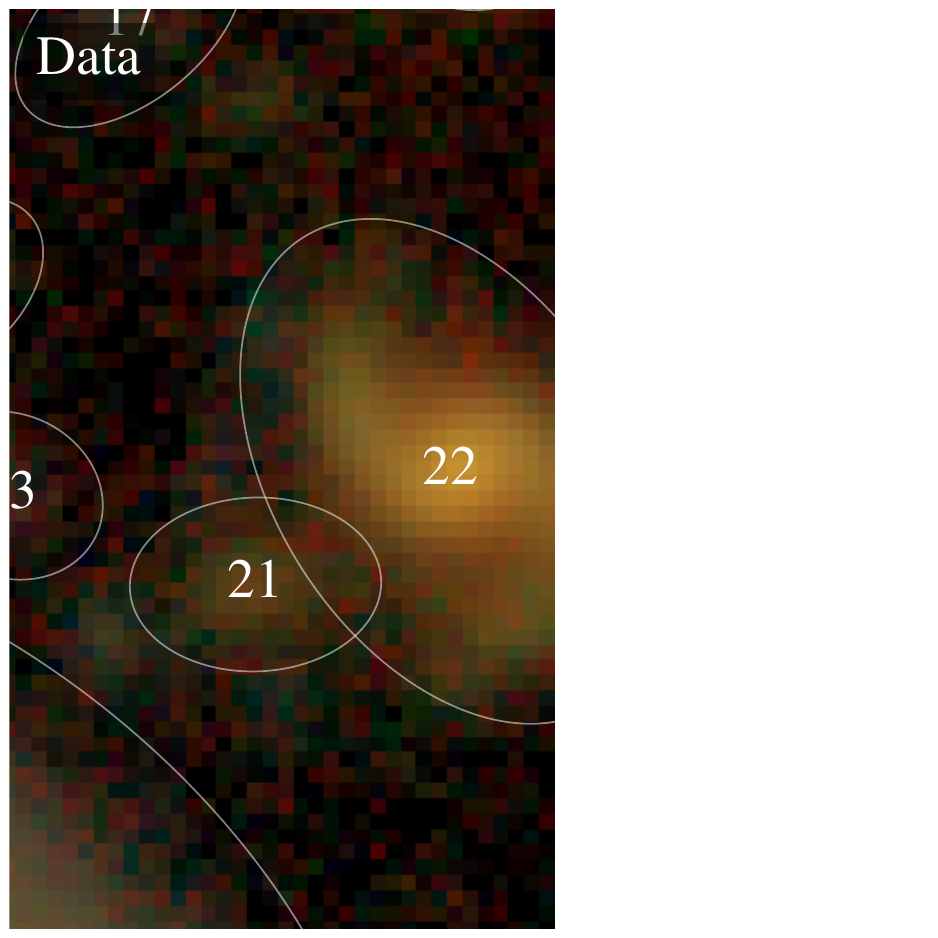}
    \includegraphics[width=.247\linewidth]{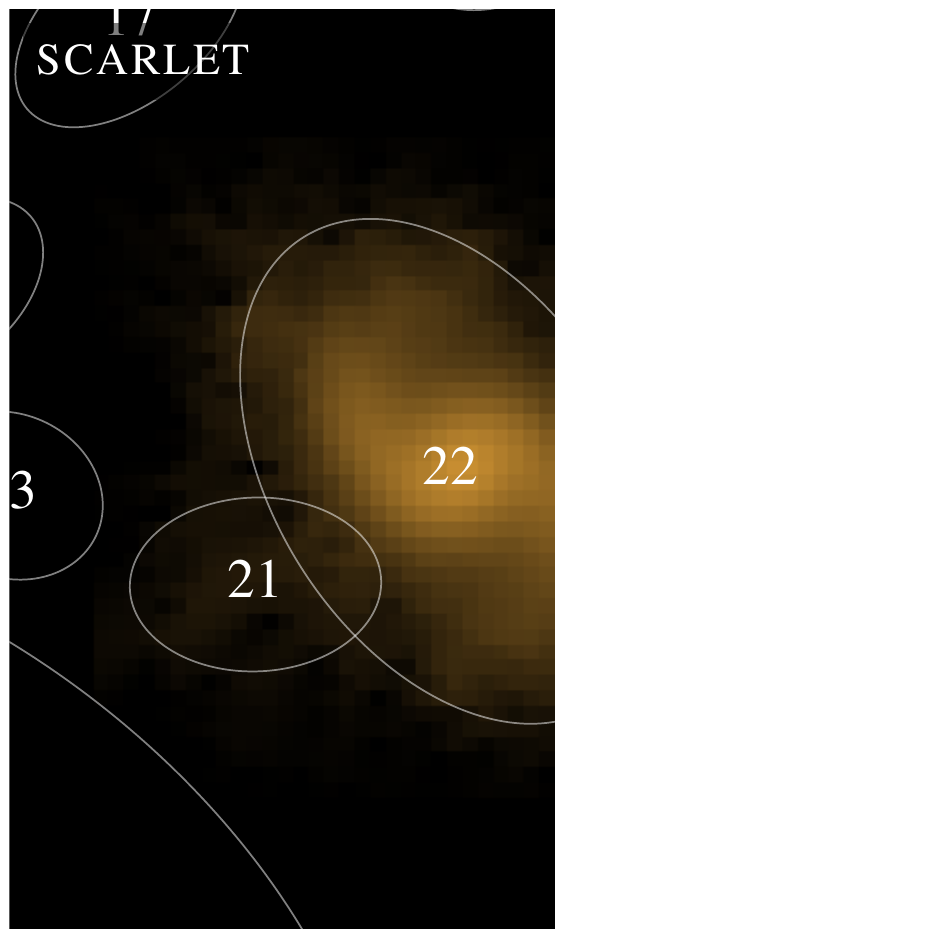}
    \includegraphics[width=.247\linewidth]{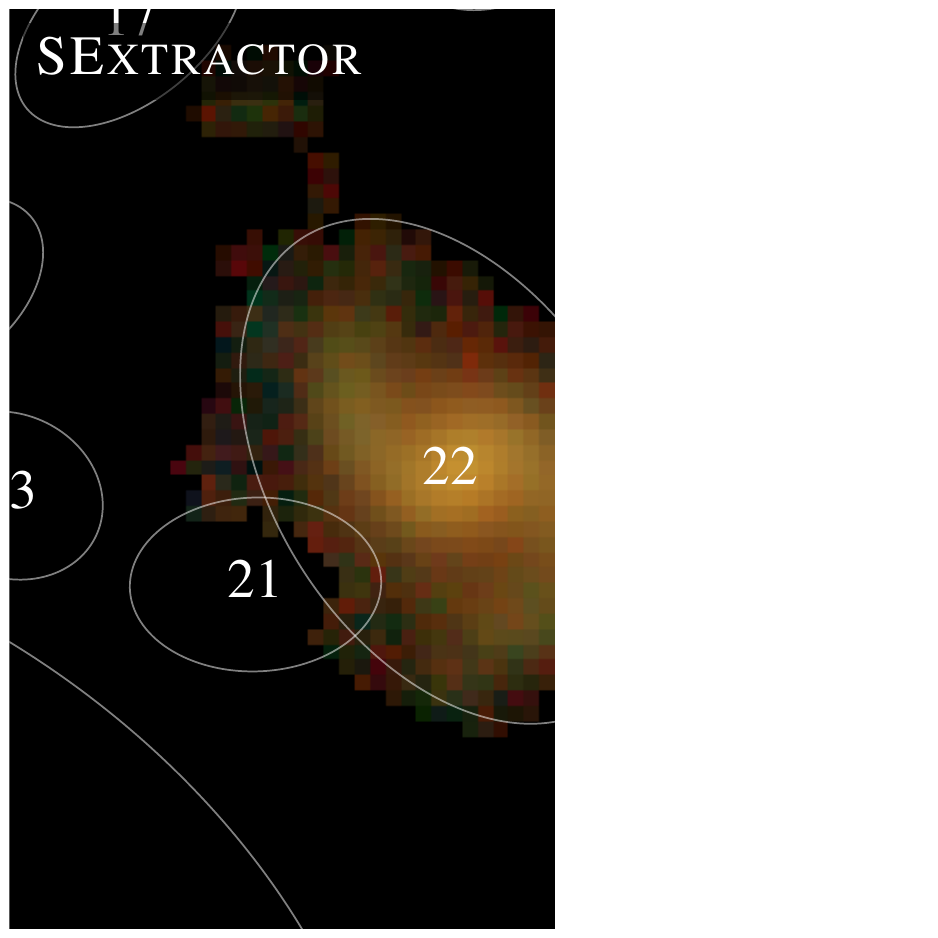}
    \includegraphics[width=.247\linewidth]{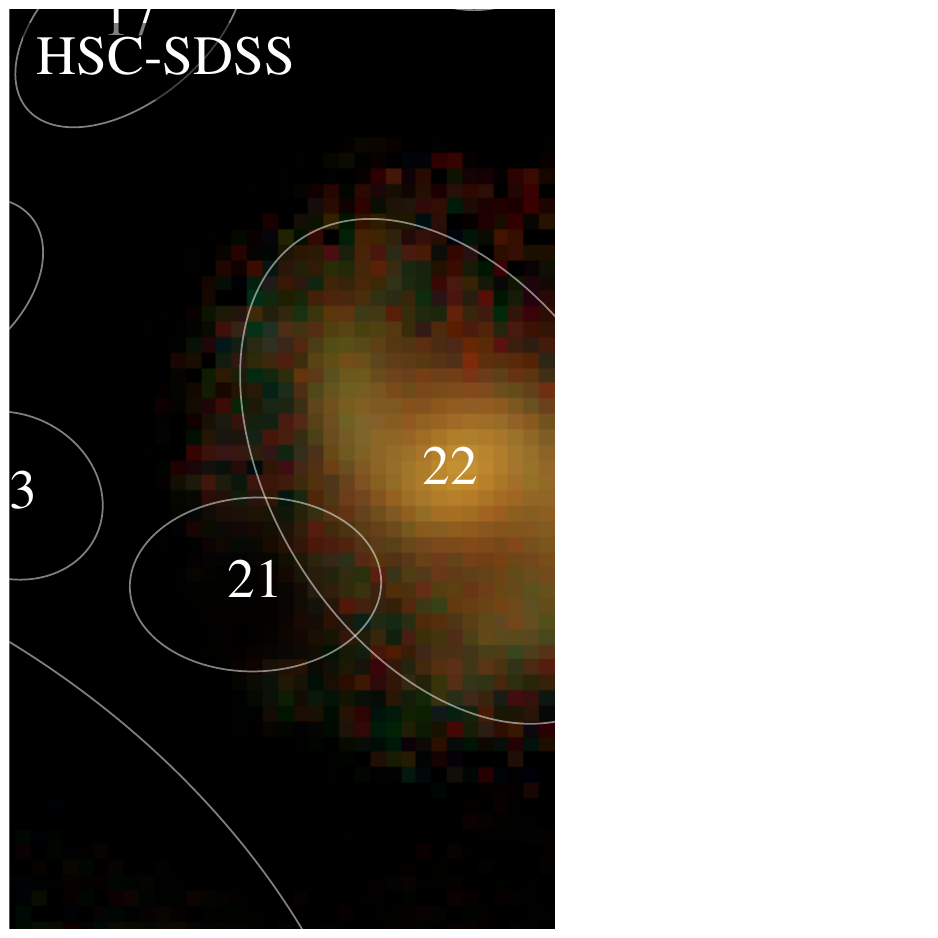}\\
    \caption{Deblender comparison for several sources (one per row) from \autoref{fig:hsc}. \emph{1st column:} HSC $grizy$ coadd data. \emph{2nd column}: The corresponding component of the \scarlet\ model. \emph{3rd column:} Data from the 1st column, restricted to the object's pixels in the {\sc SExtractor} segmentation map of the detection image. \emph{4th column}: HSC-SDSS deblender result for the object, run independently on each band.}
    \label{fig:hsc_comparison}
\end{figure*}

We demonstrate the performance of \scarlet\ in deblending deep ground-based images from HSC.  
The test case shown in \autoref{fig:hsc} stems from the UltraDeep COSMOS coadd images of the HSC public data release \citep{Aihara2017}.
The data are taken in filters {\it grizy} to a 5$\sigma$ point source depth of 27.4, 27.3, 27.0, 26.4, and 25.6, respectively, with average seeing of 0.74\,arcsec or better.
This example was selected because of its complex configuration with 33 detected sources, which span a wide range of morphologies, sizes, and brightnesses.

We detect the sources in a detection image, defined as the sum across all bands, using the {\sf Python} package {\sc sep}\footnote{\url{https://github.com/kbarbary/sep}} \citep{Barbary16.1}, which implements the functionality of {\sc SExtractor}.
The detection threshold was set at $1.2\sigma$ above the background, with very sensitive deblending parameters {\tt DEBLEND\_THRESH=64} and {\tt DEBLEND\_CONT=3e-4}; these settings were chosen to detect the majority of sources in this particular example, without over-deblending the larger galaxies in the image.

\subsubsection{Results of \scarlet}
We specify the CMF model in the simplest form, namely with single-component sources, which means the model cannot describe color variations within a galaxy.
We initialize the component morphologies $\tS_k$ by constructing symmetric, monotonic approximations to the region in the detection image around each identified peak, very similar to the approach the SDSS {\it Photo} deblender employs.
The initial SEDs $\tA_k$ are set to the color of the data averaged with the spatial weight function $\tS_k$.
We perform 200 iterations of \autoref{alg:bsdmm}, using inverse per-pixel variances as weights, but ignoring any noise correlation from resampling on the coadd image.
We also ignore the relatively small PSF variation between bands in the example, thereby performing the source separation in the convolved frame.

As one can see in the central panel of \autoref{fig:hsc}, most features of the data are represented well by the \scarlet\ model, despite the limited flexibility of one-component sources.
Furthermore, there are two resolved spiral galaxies (IDs 19 and 22), which in detail violate our assumptions of symmetric and monotonic behavior.
Nonetheless, the residuals in the right panel of \autoref{fig:hsc} indicate that this simple model can describe most of the source morphologies.
Moreover, the residuals clearly reveal that our model could be extended by adding several components to account for previously undetected localized emission, either from unrelated sources (such as the red clumps in the regions of IDs 19 and 25) or distinct stellar populations in detected sources (as in the centers of IDs 24, 29, and 33).

We leave the model refinement based on the analysis of the residuals to future work, and instead compare the deblending results of \scarlet, the HSC-SDSS deblender, and {\sc SExtractor} for several sources in \autoref{fig:hsc_comparison}.
Because {\sc SExtractor} does not attempt to separate overlapping sources (it merely seeks to find a suitable boundary between them), we show the multi-band data for the pixels belonging to the respective objects according {\sc SExtractor}'s segmentation map of the detection image.

\scarlet\ performs well on the two large spiral galaxies (IDs 19 and 22), recovering a close approximation of their shape, despite the symmetry and monotonicity constraints. 
While there will be cases where those constraints are too restrictive, e.g. for tightly wound spirals, these two examples demonstrate that galaxies with a modest amount of internal structure can be modeled well, even with a single component.
The combination of constraints is useful for dense groups of objects, e.g. around IDs 20 or 6, where e.g. a symmetry constraint alone would not suffice.
ID 6 is well modeled despite being rather faint because the simultaneous use of all available bands helps determine a best-fitting morphology, which in turn is used as a matched filter to determine the SED.
However, in IDs 20 or 25 we can see that this process can fail when undetected sources are present in the vicinity of model components.
The optimization algorithm of \scarlet\ will reduce the residuals by connecting to undetected emission via a flat ``bridge'', which is the only option available under the monotonicity constraint.
If an undetected source has a different SED, a single-component source has no other option than to adopt the flux-weighted mean SED, which is clearly visible in ID 25.
An $\ell_0$ sparsity penalty (cf. \autoref{sec:sparsity}) could suppress those bridges, however its use is problematic because it would bias low the recovery of total fluxes.

\subsubsection{Comparison to other deblenders}
{\sc SExtractor} makes no such structural assumptions, but the limitations of the ``cookie cutter'' approach clearly show with the large spirals IDs 19 and 22, as well as their neighbors, e.g. ID 20.
As a result, most objects are too compact, which will lead to low biases on total fluxes and sizes.
{\sc SExtractor} is also prone to merging different sources together, even when they are discontinuous in the segmentation map (IDs 19, 22, 25).

The HSC-SDSS deblender performs surprisingly badly on ID 19 and, by extension, on its neighbors.
Because the pixel fluxes are exactly preserved, a failure in one source has strong effects on others, e.g. IDs 20 and 25 pick up the flux from the outskirts of 19.
However, the addition of monotonicity, which was not present in the SDSS {\it Photo} implementation by Lupton (in prep.), helped in the three-in-a-row case of ID 6.

As a summary, we find \scarlet\ to perform well on a complex blended scene of deep HSC imaging, despite its simplifying assumptions of single-component sources and the rigidity of the symmetry and monotonicity constraints.
It simultaneously models all sources and is automatically consistent across all bands.
A conceptual limitation remains, namely the sensitivity to undetected sources or multiple stellar populations within detected sources, which could be mitigated by adding or increasing penalties such as a strong $\ell_0$ sparsity that restricts the support of each component to a minimum number of pixels.
On the other hand, an iterative sequence that adds components at the location of significant and coherently colored residuals appears viable and could substantially augment traditional detection methods.

\subsection{Simulations}
\label{sec:sims}

As the problem of deblending astronomical scenes is interrelated with detecting all relevant sources,
we seek to decouple these two tasks with the help of simulations where we know the number and location of each source.
The simulations also allow us to determine the accuracy of the recovered source properties, namely total fluxes, SEDs, and morphologies.
We will first test how well the assumptions of monotonicity and symmetry allow us to recover complex morphologies of isolated sources (\autoref{sec:morphology_sims}) and then how that recovery is affected when multiple overlapping objects are present in each scene (\autoref{sec:blending_sims}).
Both sets of simulations are created with the {\sc galsim} software package \citep{Rowe2015}.

As metrics to quantify the fidelity of the source modeling and deblending, we determine the fluxes, SEDs, and per-pixel morphologies of deblender results.
For the total flux, which for \scarlet\ is given by $\int d^2x\, \tS_k$, we simply compute the fractional error compared to the input model.
For SEDs and morphologies we compute the correlation coefficient between the true and the observed vectors $v$,
\begin{equation}
    \xi_v = \frac{v_\mathrm{true}\cdot v_\mathrm{obs}}{\sqrt{v_\mathrm{true}\cdot v_\mathrm{true}}\cdot \sqrt{v_\mathrm{obs} \cdot v_\mathrm{obs}}}.
    \label{eq:correlation}
\end{equation}
The coefficients $\xi_{\tA_k}$ and  $\xi_{\tS_k}$ measure how well a deblender recovers colors and shapes (including intensity) of component $k$, respectively.
We consider these fidelity measures as a function of true source flux and source ``blendedness'' \citep{Bosch17}
\begin{equation}
\label{eq:blendedness}
    \beta = 1-\frac{\tS_k\cdot \tS_k}{\tS_{\textrm{all}} \cdot \tS_k},
\end{equation}
where $\tS_\mathrm{all} = \sum_k \tS_k$, expecting that good results will be harder to achieve for faint sources in very crowded areas.

\subsubsection{Individual galaxies}
\label{sec:morphology_sims}

\begin{figure}[h]
   \includegraphics[width=\linewidth]{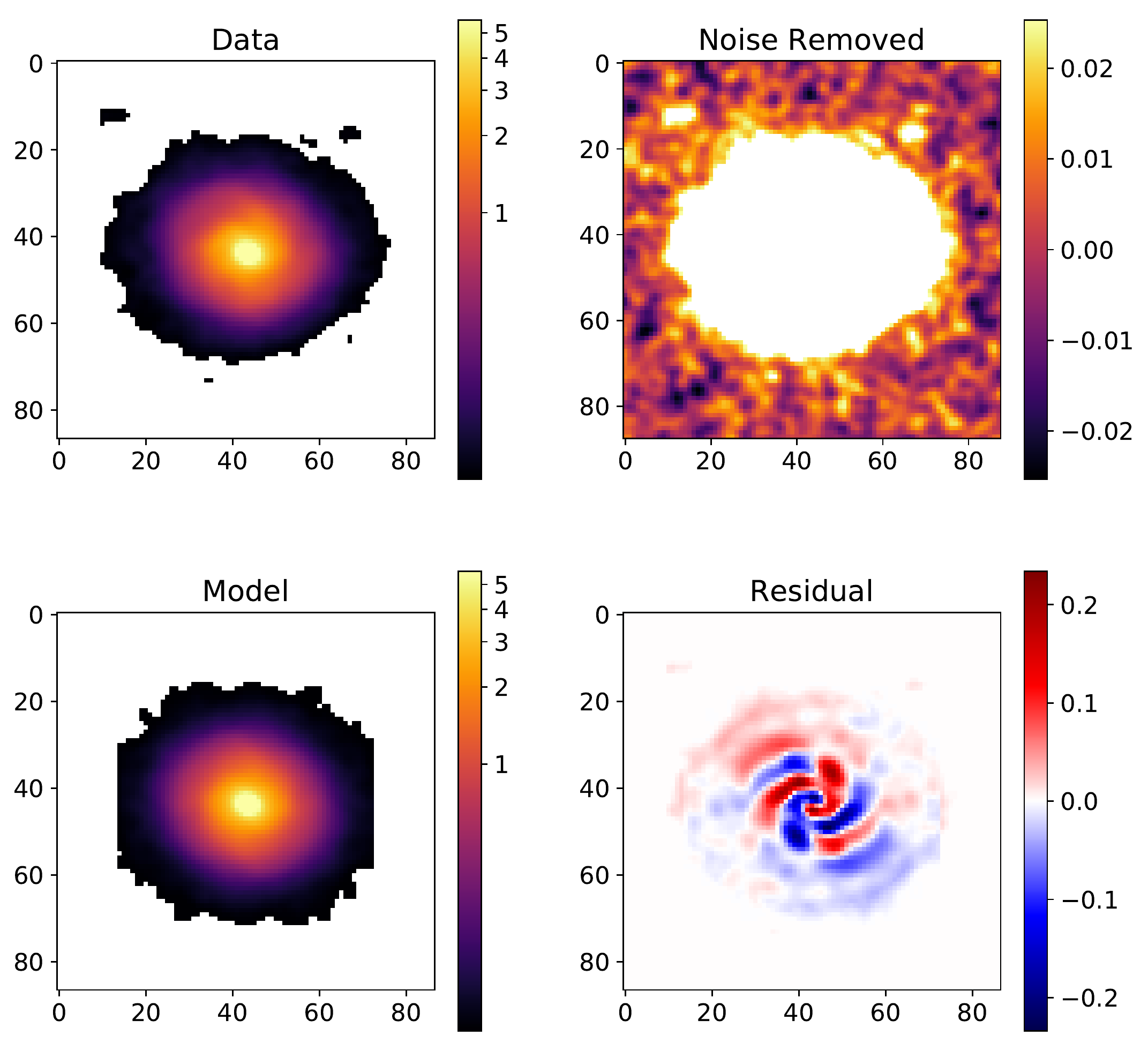}
    \caption{An example galaxy from the COSMOS {\it HST} dataset in {\sc galsim}. 
    \emph{Top left:} The ``truth'' image, defined by an original {\it HST} image from which we removed the noise in the galaxy outskirts. 
    \emph{Top right:} The difference between the original {\it HST} image and the ``truth''.
    \emph{Bottom left:} The model generated by \scarlet\ with symmetry and monotonicity constraints.
    \emph{Bottom right:} The residuals (truth $-$ model) reveal spiral structure that is missing in the \scarlet\ model.     
    The panels on the left use a $\textrm{sinh}^{-1}$ scaling. 
    }
    \label{fig:real_morph}
\end{figure}

We start by investigating how well \scarlet\ can reproduce realistic galaxy morphologies.
We use a set of galaxies from the COSMOS {\it HST} sample described in \citet{Mandelbaum2012} that are available with {\sc galsim}\footnote{\url{https://github.com/GalSim-developers/GalSim/wiki/RealGalaxy\%20Data}}.
From the full sample of 87,000 galaxies with $i<25.2$, we further select those with total fluxes between 300 and 2,000 counts that pass most stringent ``marginal'' cuts.
The bright cutoff was chose to remove galaxies with visible blending: many of the brightest galaxies have faint background galaxies or foreground stars in their wings.
As we seek to test the recovery of \emph{individual} sources here, those interlopers are contaminants that reduce the correlation measure in \autoref{eq:correlation}.
They could easily be fit with \scarlet\ if their existence and positions were known, but for the simulations here we want to assume that we know the position of every individual source.
The lower flux cutoff was imposed to eliminate faint sources, whose outskirts are dominated by correlated noise from the original {\it HST} observations.
To further reduce the effect of correlated noise, we estimate the noise level over the postage stamp and subtract the corresponding value from the entire image, setting all negative pixel values to zero.
While this procedure is likely to cut off some flux from the wings of the galaxy, the bulk of the galaxy and especially its internal structure is preserved.
The remaining sample comprises 799 galaxies.
An example galaxy from this procedure is shown in \autoref{fig:real_morph}.

\begin{figure}[h]
   \includegraphics[width=\linewidth]{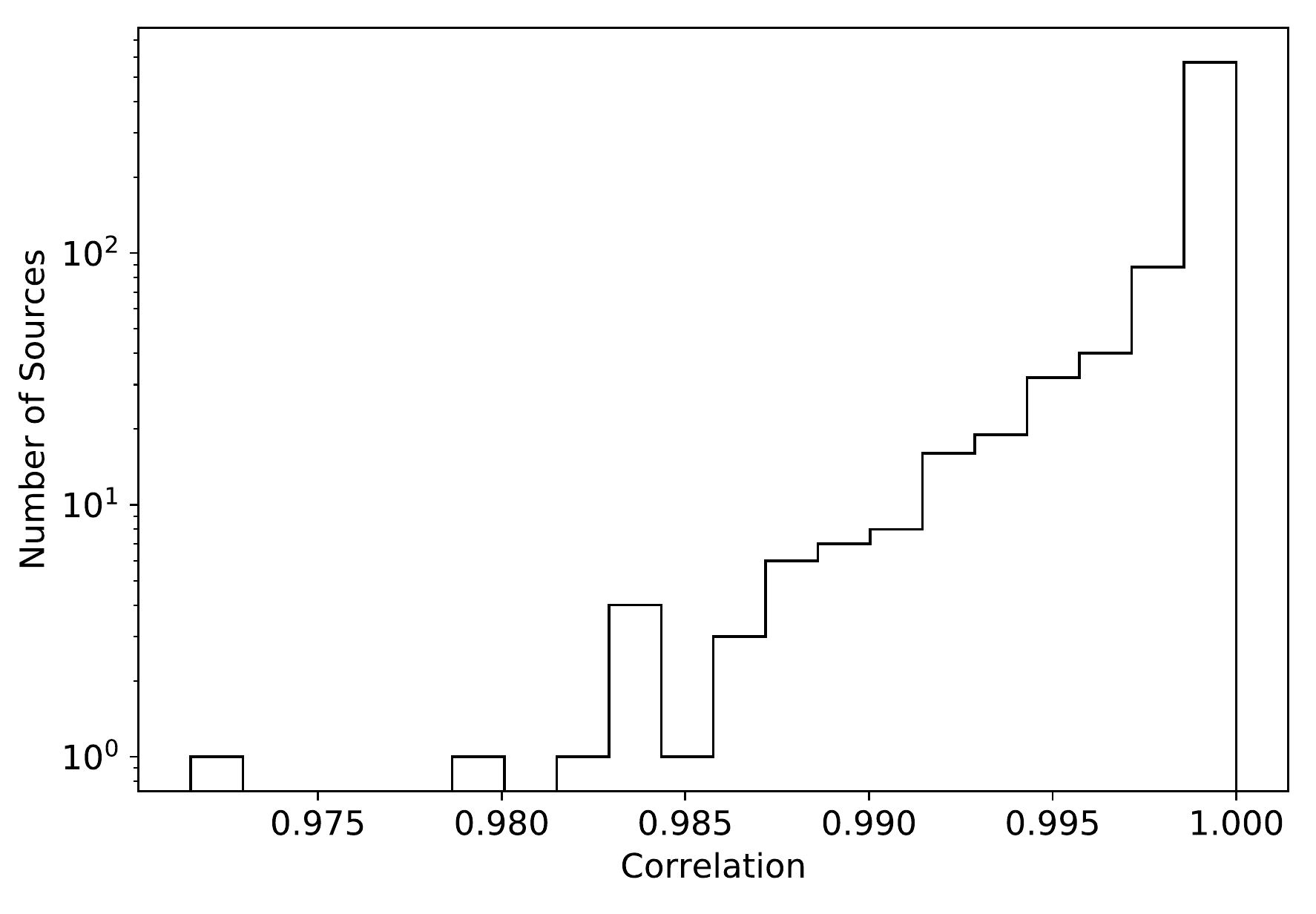}
    \caption{Morphology correlation coefficient (\autoref{eq:correlation}) of \scarlet\ models for 799 isolated galaxies from the COSMOS {\it HST} sample in {\sc galsim}.}
    \label{fig:real_morph_correlation}
\end{figure}

We run \scarlet\ on single-band images of all 799 galaxies. The pixel resolution is set to a typical ground-based value of 0.2\,arcsec; no noise is added to the image.
We show the resulting model and the residuals for the example galaxy in the bottom panels of \autoref{fig:real_morph}.
For reference, its total flux error is 0.08\%, its morphology correlation coefficient $\xi_\tS=0.999$.
The RMS of the total flux error for the entire sample is 0.39\%, which increases for fainter sources, suggesting that some correlated noise still made it into the input images and is suppressed by the monotonicity constraint.
The morphology correlation $\xi_\tS$ is shown as a cumulative distribution function in \autoref{fig:real_morph_correlation}.
These results demonstrate that for realistic galaxies, \scarlet\ recovers morphologies with a very strong correlation to the input morphologies despite the constraints of symmetry and monotonicity; consequently, the estimation of total fluxes is also excellent.

\subsubsection{Blended scenes}
\label{sec:blending_sims}

We now create a set of 10,000 blended scenes in the LSST filters {\it ugrizy}, with widths and heights varying between 30 and 70 pixels, comprising on average a mix of 20\% stars and 80\% galaxies, with a common PSF across all bands.
The stellar sources have SEDs taken from the \citet{Pickles1998} templates for O5V, B5V, A5V, F5V, K5V, and M5V stars, and each stellar morphology is defined as a 2D Gaussian with $\sigma=10^{-9}$\, pix that is convolved with the PSF in each band.
Galaxy sources have SEDs based on four empirical \citep[CWW]{Coleman1980} templates available in {\sc galsim}, randomly perturbed to generate a total of 25 different galaxy SED templates.
Their redshifts are randomly chosen between 0 and 2.1.
For the galaxy morphologies we use the same 799 galaxies selected in \autoref{sec:morphology_sims}.
All sources are inserted into the scenes with a power-law distribution in radius from the center, resulting in strongly crowded blends in the middle of the images.
Uncorrelated Poisson noise is added to the images, using a sky background brightness at the mean noise level of the {\it HST} images, so that we can largely hide the truncation of the input galaxy morphologies (cf. \autoref{fig:real_morph}).

\begin{figure}[t!]
    \includegraphics[width=\linewidth]{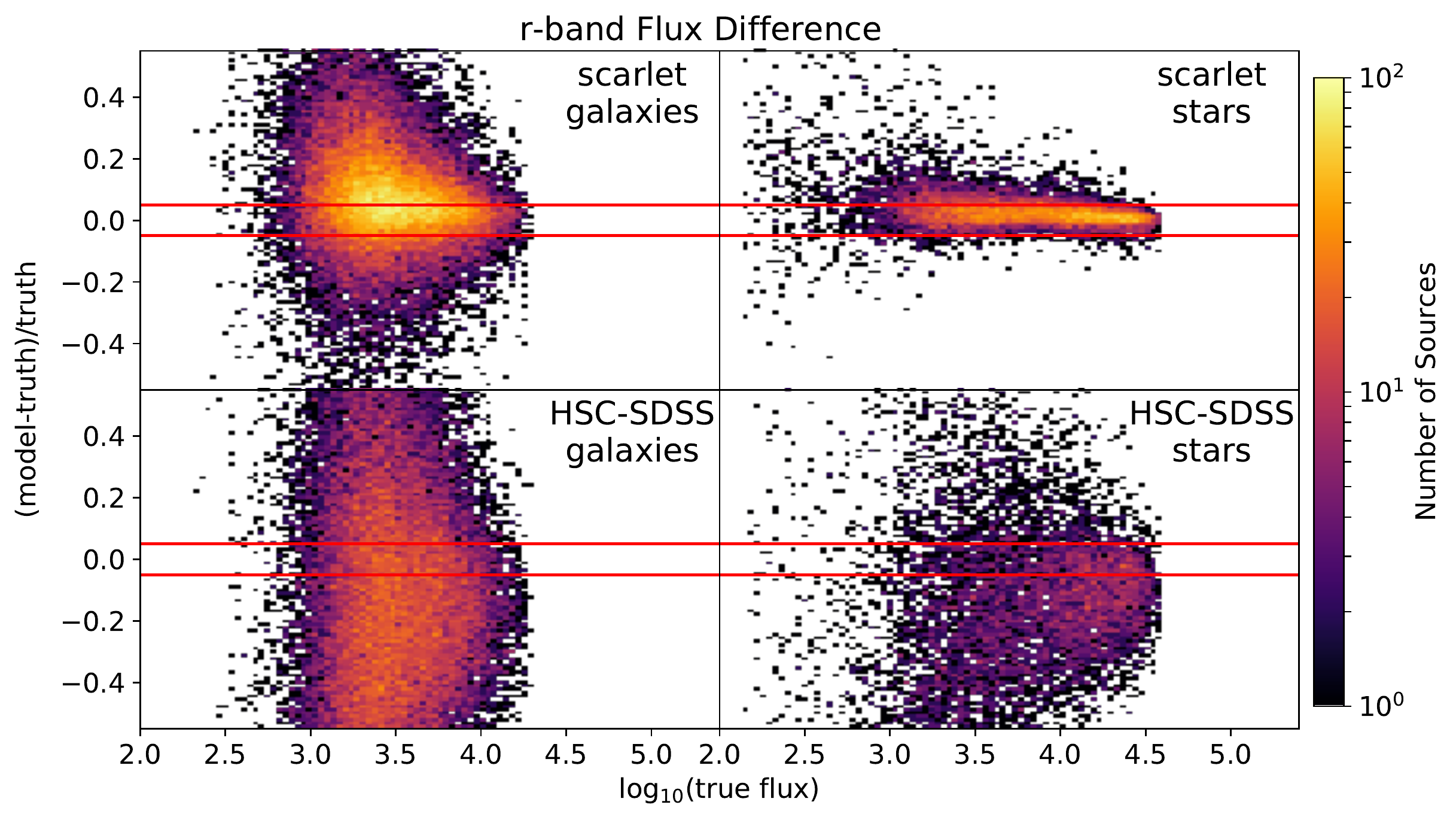}
    \caption{Comparison of the relative error of total fluxes in the $g$ between \scarlet\ (top) and the HSC-SDSS deblender (bottom), for galaxies (left) and stars (right).
    For reference, the red horizontal lines show 5\% relative errors.
    We only show the results of the $r$-band as the other bands look qualitatively similar.
    }
    \label{fig:flux}
 \end{figure}

\begin{figure}[ht]
   \includegraphics[width=\linewidth]{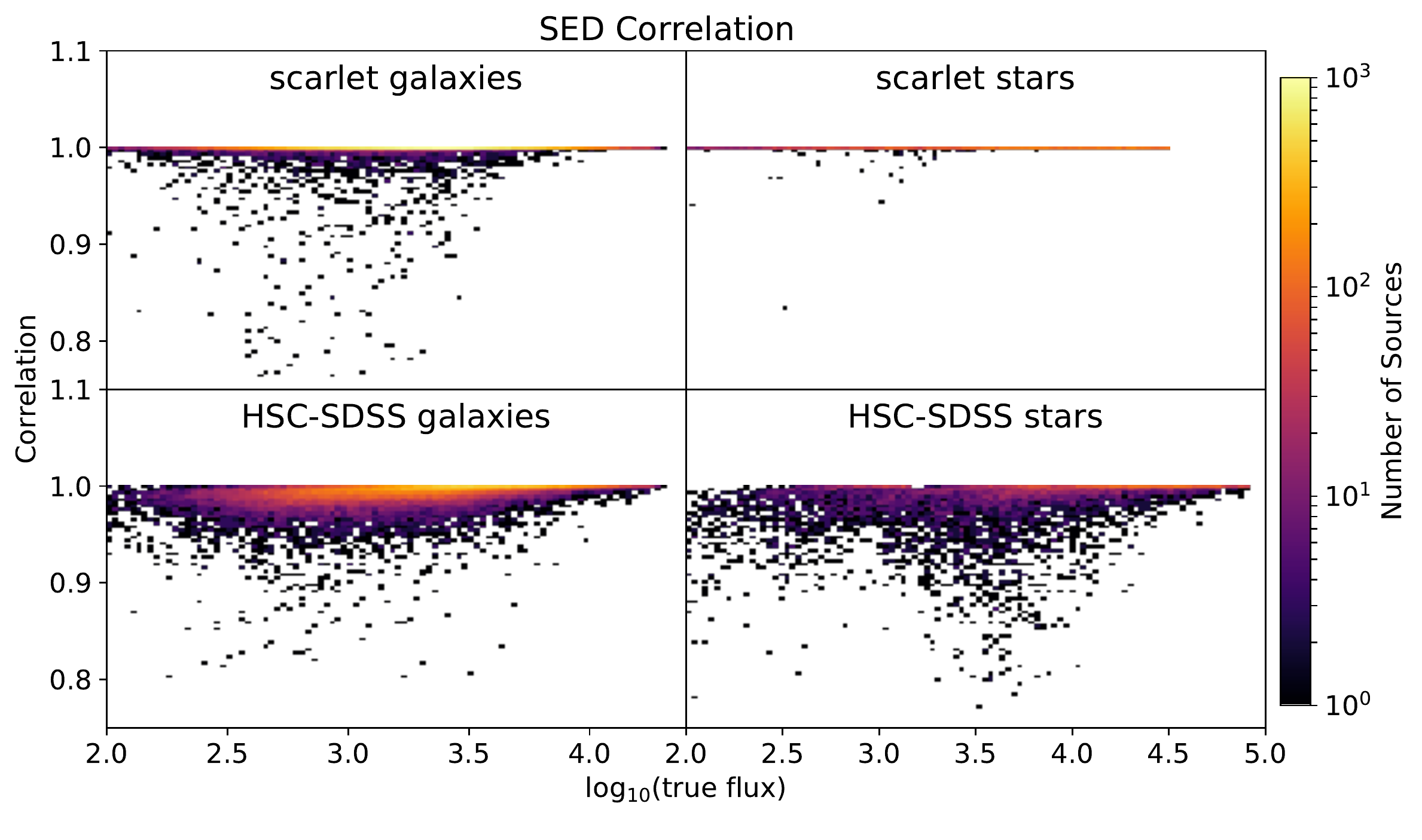}
   \includegraphics[width=\linewidth]{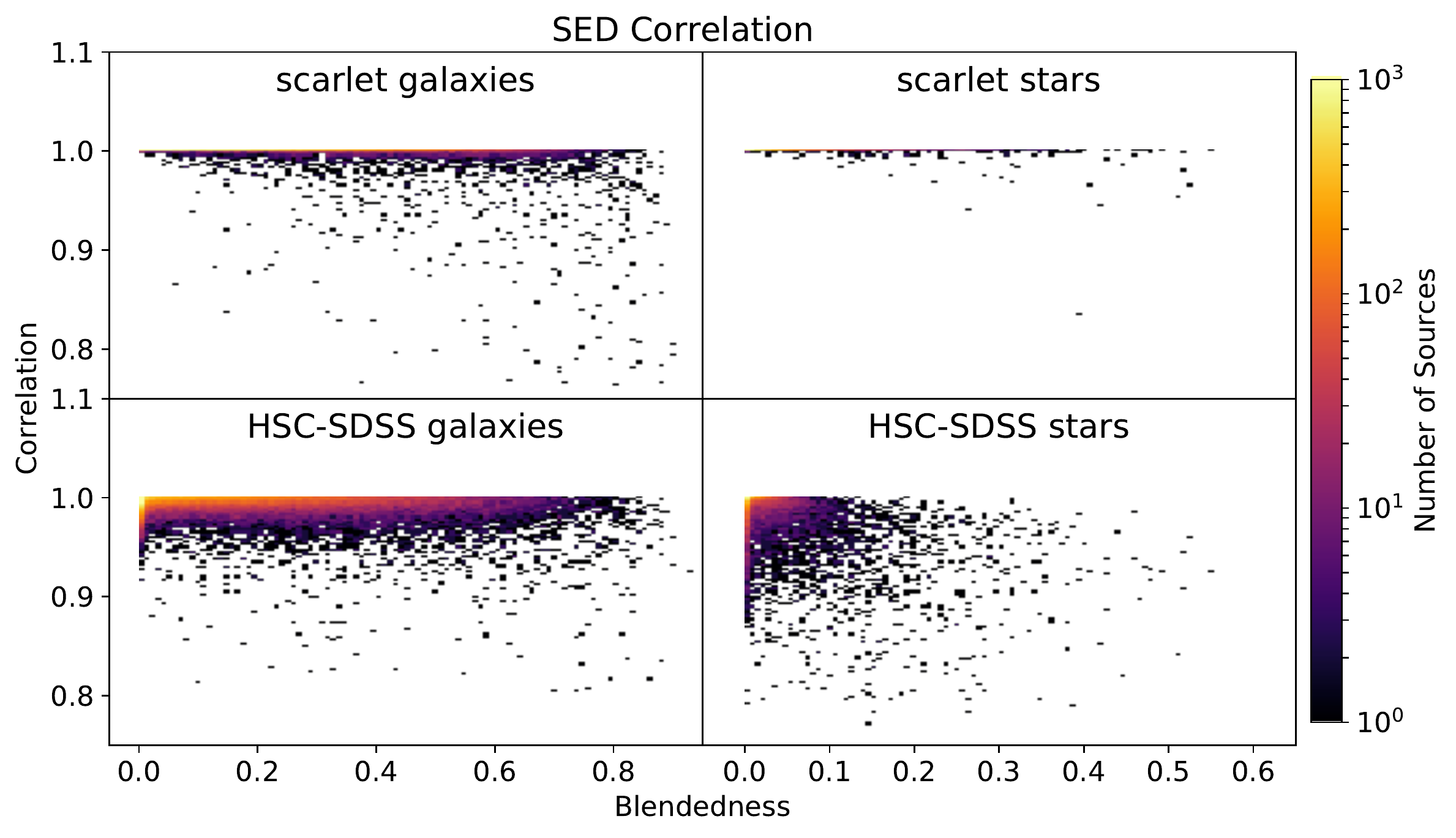}\\
   \caption{Correlation, as defined by \autoref{eq:correlation}, between true SEDs and the SEDs from the HSC-SDSS deblender or \scarlet\, as a function of true source flux (top four panels) and blendedness (\autoref{eq:blendedness}, bottom four panels.)}
    \label{fig:sed}
\end{figure}

\begin{figure}[h]
   \includegraphics[width=\linewidth]{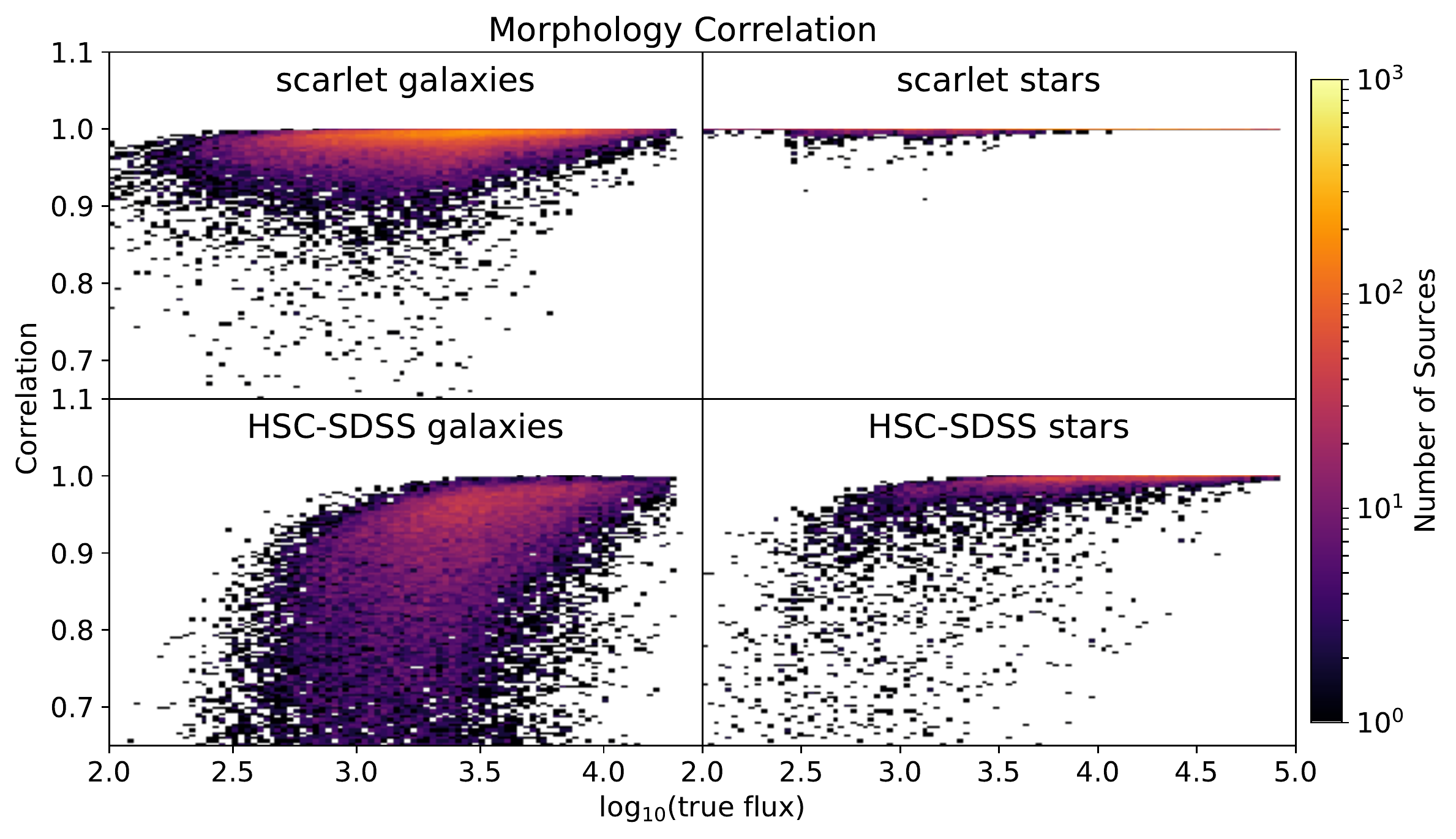}
   \includegraphics[width=\linewidth]{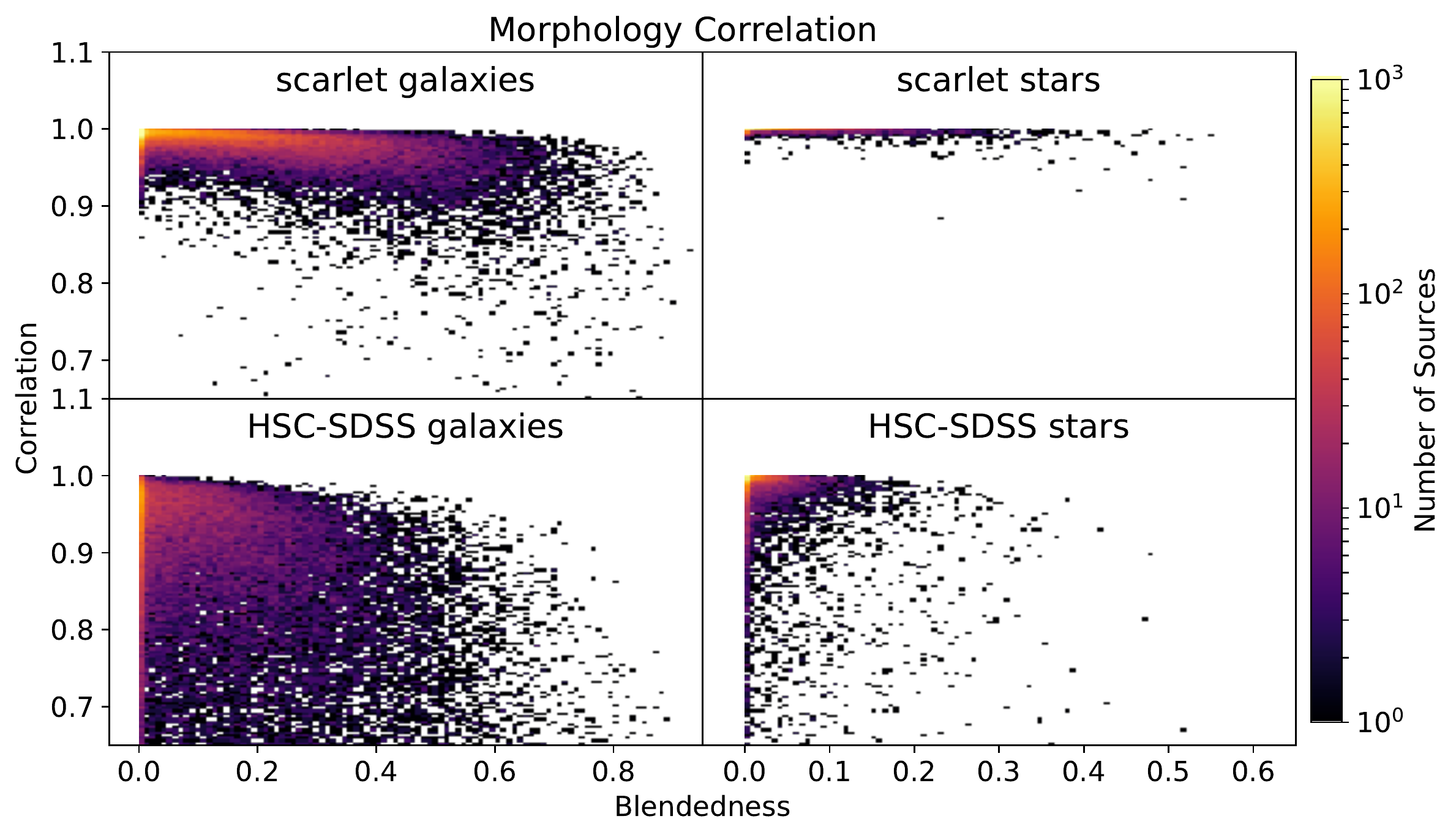}\\
    \caption{Same as \autoref{fig:sed} for the correlation between the true source morphology and the morphologies from the HSC-SDSS deblender or \scarlet. For HSC-SDSS, the morphology is taken from the $g$-band; other bands look similar.}
    \label{fig:morph}
\end{figure}

We run \scarlet\ and the current HSC-SDSS deblender and separate the results for stars and galaxies assuming that larger average brightness and more compact shape should make stars easier to deblend than galaxies.
Indeed, we find in \autoref{fig:flux} that total flux errors are smaller for stars than for galaxies and that \scarlet\ substantially outperforms the HSC-SDSS deblender.
These two trends continue with the correlation coefficients of SEDs in \autoref{fig:sed} and of morphologies in \autoref{fig:morph}.
We understand the remarkable difference between these two deblending strategies, which have very similar constraints on the source morphologies, as a result of fitting a joint model across all bands (\scarlet) or independent models for each band (HSC-SDSS).
Our strategy is superior in these test cases because we combine images of all bands, resulting in a morphology that is driven by the band with the highest significance, which is in turn used as an optimal filter to determine the source SED. 

We emphasize that these results are achieved with single-component, symmetric, and monotonic source models, even though the true morphologies in the simulations exhibit more complex structure.
There will be cases in which those constraints lead to substantial errors, but we believe that they are now demonstrably justified as powerful default configurations for deblending in ground-based wide-field surveys.

We finally note that the runtime of \scarlet\ in our tests is about 10\,ms per band and source, driven almost entirely by the cost to evaluate the likelihood gradients in \autoref{eq:full_gradients}.
The computational cost increases substantially if the PSF convolution cannot be expressed efficiently.
While slower than e.g. the HSC-SDSS deblender, \scarlet\ only has to be run once over all images and directly yields measurements of flux, SED, and morphology of the sources.
We fully expect source extraction with \scarlet\ to be a computationally expensive but still feasible task for deep optical surveys.

\subsection{Extended Emission Line Regions of AGN}
\label{sec:jet}

\begin{figure*}[ht]
    \includegraphics[width=.247\linewidth]{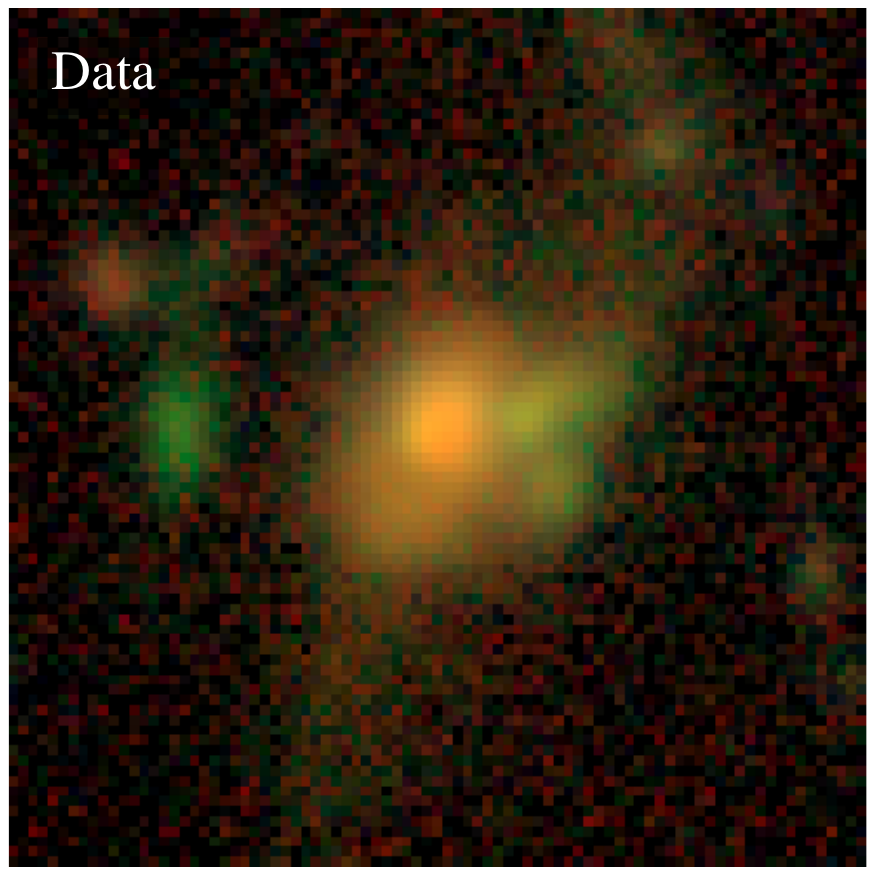}
    \includegraphics[width=.247\linewidth]{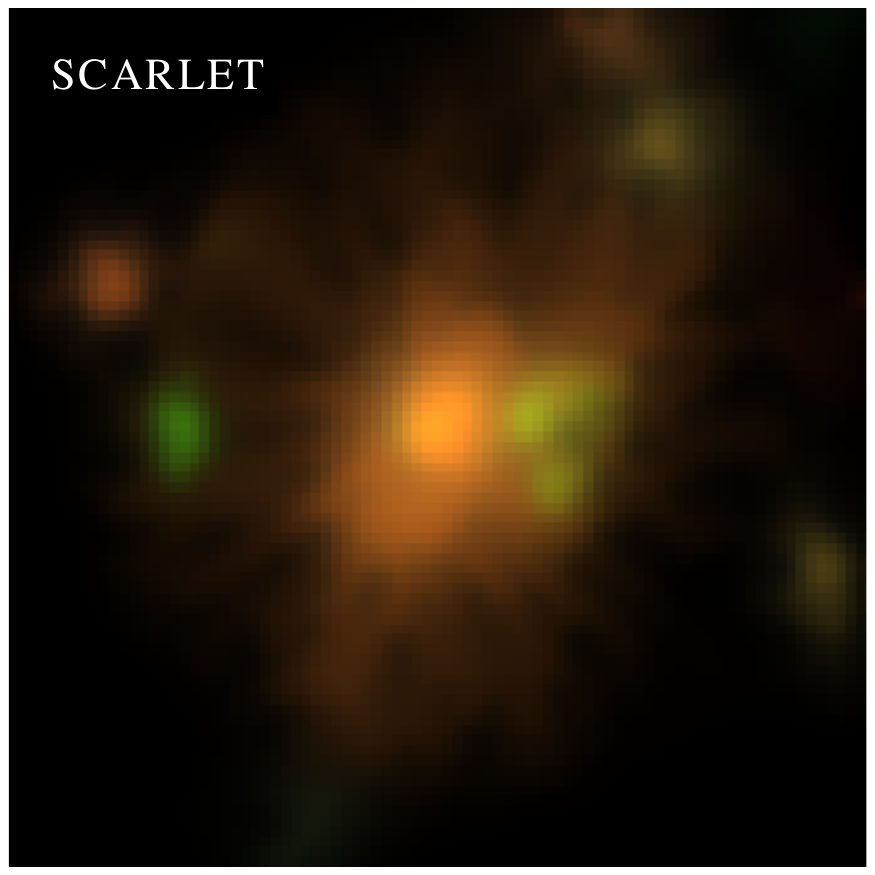}
    \includegraphics[width=.247\linewidth]{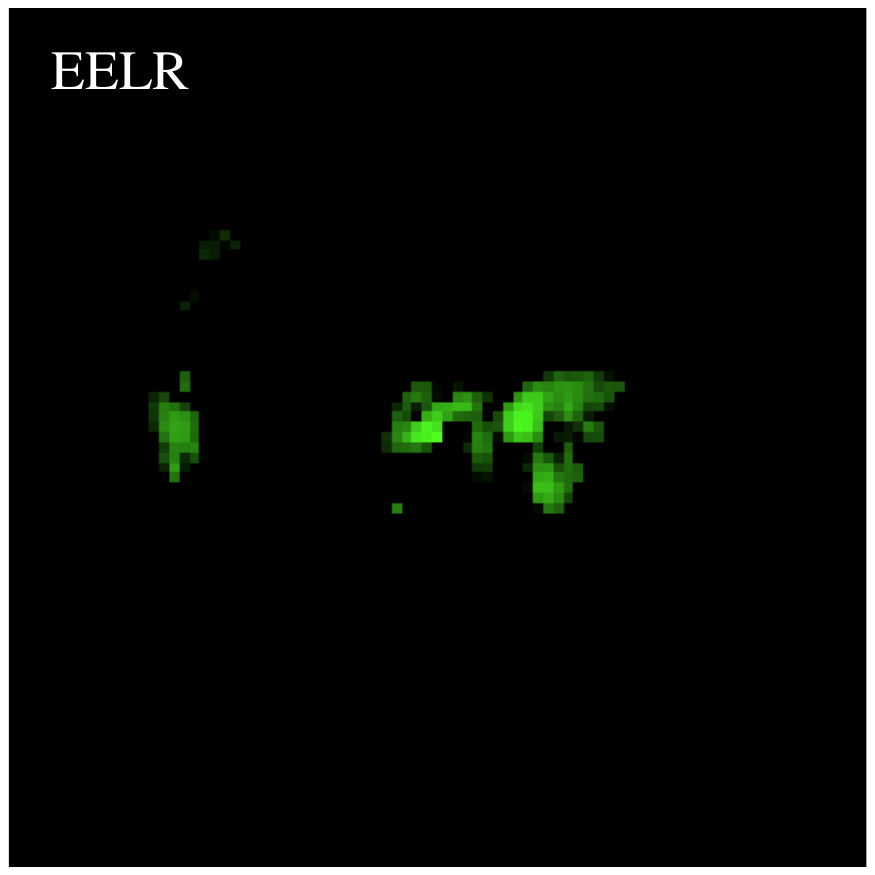}
    \includegraphics[width=.247\linewidth]{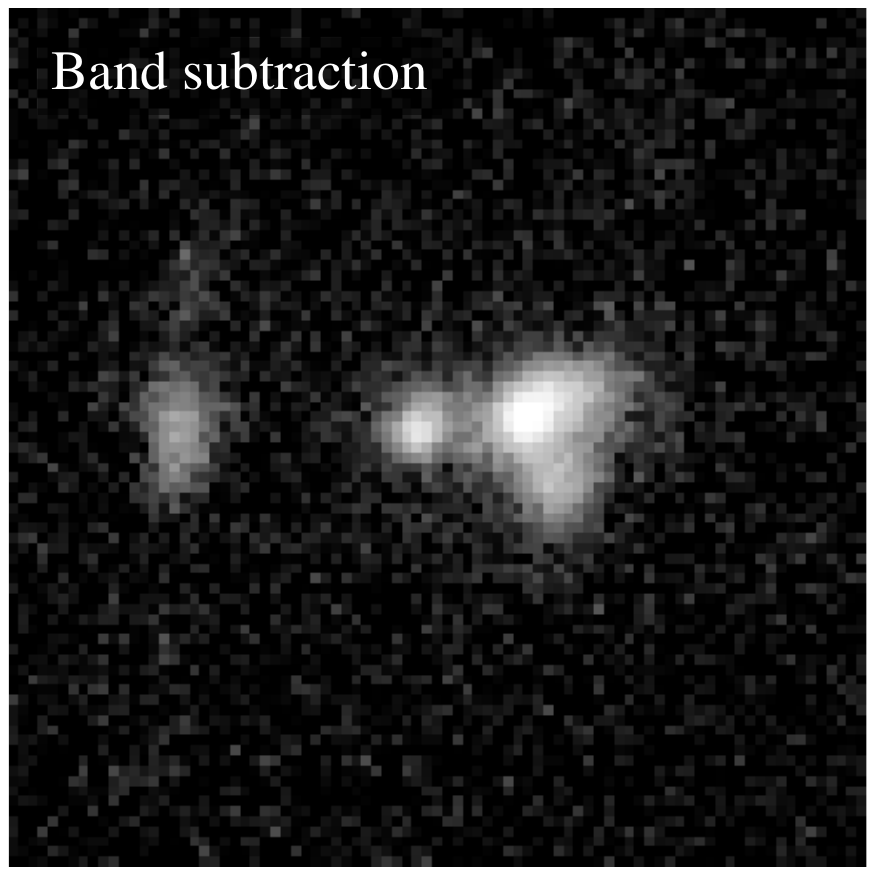}\\
    \caption{\emph{1st panel:} False-color image of $grizy$ coadd images of \object{SDSSJ023106-034513} from the HSC Wide data release, shown with an arcsinh stretch. \emph{2nd panel:} Convolved \scarlet\ model of the scene with a two-component monotonic source for the host galaxy, a free-form source for the EELR, and monotonic and symmetric sources for other detections in the scene. \emph{3rd panel:} Partially deconvolved EELR component. \emph{4th panel:} Result of the band subtraction $i-z$ from \citet{Sun2018}, which subtracts the galactic continuum to isolates the EELR emission. }
    \label{fig:agn_jet}
\end{figure*}

To further demonstrate the capabilities of \scarlet\ as a flexible component separation framework, we analyze HSC Wide coadd images in $grizy$ bands of galaxies with extended emission-line regions (EELRs) created by active galactic nuclei (AGN). 
Resulting from AGN radiation on inhomogeneous inter-stellar medium, the morphology of EELR is often complex, asymmetric, and non-monotonic. 
The strong emission lines, e.g., [OIII]$\lambda$5007, emitted can be detected and resolved by broadband imaging as components that may spatially overlap with but have colors distinct from their host galaxies \citep[e.g.,][]{Lintott2009,Keel2012}. 

The particular example shown in \autoref{fig:agn_jet}, \object{SDSS J023106-034513}, is taken from the EELR sample of \citet{Sun2018}, who selected EELR candidates from spectroscopically identified type 2 obscured AGNs. 
With a high luminosity of $L_\mathrm{bol} = 10^{45.7}$ erg s$^{-1}$, this AGN creates extended [OIII]$\lambda$5007 emission with a diameter of 60 kpc that is captured in the HSC $i$-band image, see \autoref{fig:agn_jet}. 

We seek to separate the EELR from its host galaxy with \scarlet.
We first detect all sources in the image, and then assign them to either host, EELR, or one of 3 likely unrelated interlopers.
We constrain the morphology of the interlopers with our default combination of non-negativity, monotonicity and symmetry.
The host galaxy does not appear symmetric, so we remove that constraint\footnote{In experiments, we found that the host is better modeled by two co-centered components, similar to a bulge-disk model.},
while the EELR has no constraints other than non-negativity.

To compensate for the weakening of morphological constraints, we can utilize our knowledge of the EELR broadband SED based on the emission lines in the observed SDSS spectrum.
Similarly, by assuming that the host galaxy is responsible for the continuum spectrum without emission lines, we can also predict its broadband SED.
The proximal operator for $\tens{A}_k$ is thus simply a projection onto the precomputed SEDs of EELR or host and mute for any of the interlopers:
\begin{equation}
\prox_\mathrm{AGN}(\tens{A}_k) = \begin{cases}
\tens{A}_\mathrm{EELR} & \text{if}\ k = k_\mathrm{EELR}\\
\tens{A}_\mathrm{Host} & \text{if}\ k = k_\mathrm{Host}\\
\tens{A}_k & \text{else.}
\end{cases}
\end{equation}
We use inverse per-pixel variances as weights and match the observed PSFs between bands by deconvolving to an effective resolution of 0.35 arcsec, using the formalism of \autoref{sec:psf}.
We show the converged, convolved \scarlet\ model in the second panel of \autoref{fig:agn_jet} and the partially deconvolved EELR component in the third panel.
For comparison, in the forth panel, we also show the EELR image obtained via image continuum subtraction from \citet{Sun2018}. 
In this approach, to isolate the EELR signal, the galaxy continuum in the $i$-band is removed by subtracting the (scaled) $z$-band image, which contains only the continuum but not line emissions. 
We can see that the \scarlet\ model of the EELR strongly resembles the continuum-subtraction result, while revealing smaller spatial features of the EELR.
A more detailed application of \scarlet\ to EELR extraction is planned for future work.

\section{Summary and Outlook}
\label{sec:conclusions}

The employment of multi-channel information is long known to aid source separation schemes.
It was therefore expected that multi-band imaging should be beneficial in cases when multiple sources overlap in a celestial scene.
To exploit the availability of multi-band imaging from modern sky surveys, we have developed the source separation framework \scarlet.
It uses a matrix factorization scheme, in which one matrix factor describes the amplitude of all components in each band and the other factor describes their spatial shape.
We use the proximal optimization method bSDMM presented by \citet{Moolekamp2017} to allow for an arbitrary number of constraints on each matrix factor.
The astrophysical interpretation of this model is that the scene is composed of a number of components that each describe one star or distinct stellar population.
Galaxies can be modeled as several such components to characterize e.g. bulge and disc-like features.

To aid the source separation further, especially in crowded regions with little variation in source colors (such as galaxy cluster cores), we constrain the component morphologies to be monotonic and/or symmetric with respect to the sources peak position.
If available, constraints on the SED of the sources, e.g. from stellar SED libraries or high-resolution spectra, are beneficial as well.

We have derived the analytical treatment of correlated pixel noise and of band-dependent PSF convolution, rendering \scarlet\ capable of working on coadd images with variable seeing conditions.
In essence, it performs a PSF-matched photometry measurement with an optimally chosen weight function given by the signal-to-noise-weighted mean morphology in all available bands.
Applied to an example scene from the HSC UltraDeep survey, we find \scarlet\, to robustly recover all detected sources, even if they exhibit complex morphologies.
For detailed analyses of extended galaxies one can conceive of additional constraints, such as the Fourier mode extension of {\sc Galfit} \citep{Peng10.1}.

We show with a set of dedicated simulations comprised of stars and realistic galaxy morphologies that \scarlet\ yields an accurate recovery of source fluxes, SEDs, and morphologies, even for prominent levels of blendedness.
In our tests it clearly outperforms the HSC-SDSS deblender in each of these measurements and is especially powerful for stars and marginally resolved as well as faint sources.

Because of its non-parametric nature, \scarlet\ exhibits a set of inherent advantages and drawbacks, which we summarize in \autoref{tab:pro_con}. 
In addition, due to its flexibility it is sensitive to undetected sources or additional components of detected sources.
On the other hand, the inspection of the fit residuals appears to clearly reveal their presence.
We therefore envision a scheme in which components are added at the location of significant and coherently colored residuals.
A decision whether any newly added component is part of a previously detected source or independent of all other sources will require a post-processing step that encodes an interpretation based on the nature of the astronomical scene.

When run in this fashion \scarlet\ breaks a common model of data processing in optical astronomy that consists of three distinct stages: detection, deblending, and characterization.
If components are added from an analysis of the residuals and the models are used to infer source properties, it essentially covers aspects of all three stages.
While doing so, it makes assumptions about the SEDs and morphologies of the sources as well as the noise and the PSFs in the observations, which need to be valid to produce reliable results.
It will therefore be necessary to assess the accuracy of \scarlet\ models for specific science cases.
To facilitate inference directly on \scarlet\ sources, we intend to provide an error-estimation scheme, which accurately captures component degeneracies and effects of non-differentiable constraints.

We have intentionally developed \scarlet\ such that it can utilize any constraint or prior that can be expressed as a proximal operator without having to change the algorithm itself, thus rendering it suitable to a wide range of source separation problems in image data.
We demonstrated this flexibility with an AGN jet--host galaxy separation in this work and with a hyperspectral unmixing task in \citet{Moolekamp2017}.
It is inherent to the matrix factorization approach that \scarlet\ should work well in applications where high-quality information on the spatial distribution of sources is to be combined with high-quality information on their spectra.
An example is the combination of a single high-resolution image from a space-born telescope, such as {\it HST} or the future {\it WFIRST} \citep{Spergel15.1}, with multi-band imaging of lower spatial resolution from a ground-based telescope.
We will pursue this research direction in forthcoming works.

\scarlet\ is implemented in {\sf Python} with {\sf C++} extensions, accessible through {\sf pybind11}, and is available at \url{https://github.com/fred3m/scarlet}.
Its interface is highly modular and should be easily extendible to different applications.
Contributions and requests from the astronomical community are highly encouraged.

\begin{table*}[ht]
\caption{Main advantages and drawbacks of \scarlet.}
\label{tab:pro_con}
{\small
\begin{center}
\begin{tabular}{lll}
{\bf Feature} & {\bf Advantage} & \bf{Disadvantage}\\
 \hline\hline
Matrix Factorization & Low-rank representation of celestial scenes & Not suited for correlated changes of SED \emph{and} morphology\\
\rule{0pt}{3ex}Non-parametric model & Arbitrary morphologies and SEDs & Requires critical sampling \\
& & No analytic PSF convolution\\
\rule{0pt}{3ex}Proximal gradient optimization & Wide range of priors and penalties & No analytic error propagation
\end{tabular}
\end{center}
}
\end{table*}%

\section*{Acknowledgements}
We gratefully acknowledge support from NASA grant 14-WPS14-0003.

This material is based upon work supported in part by the National Science Foundation through
Cooperative Agreement 1258333 managed by the Association of Universities for Research in Astronomy (AURA), and the Department of Energy under Contract No. DE-AC02-76SF00515 with the SLAC National
Accelerator Laboratory. Additional LSST funding comes from private donations, grants to universities,
and in-kind support from LSSTC Institutional Members.

\section*{References}

\nocite{Miyazaki2018}
\nocite{Komiyama2018}
\nocite{Furusawa2018}

\bibliography{references.bib}

\end{document}